\documentclass[letterpaper]{article} 
\usepackage{booktabs}
\usepackage{multirow}
\usepackage{listings}
\usepackage{amsmath}
\usepackage{color}
\usepackage{xcolor}
\usepackage{colortbl}
\usepackage{subfigure}
\usepackage{aaai2026}  
\usepackage{times}  
\usepackage{helvet}  
\usepackage{courier}  
\usepackage[hyphens]{url}  
\usepackage{graphicx} 
\usepackage{natbib}  
\usepackage{caption} 
\usepackage{algorithm}
\usepackage{algorithmic}

\usepackage{newfloat}
\usepackage{listings}
\DeclareCaptionStyle{ruled}{labelfont=normalfont,labelsep=colon,strut=off} 
\floatstyle{ruled}
\newfloat{listing}{tb}{lst}{}
\floatname{listing}{Listing}
\definecolor{dkgreen}{rgb}{0,0.6,0}
\definecolor{gray}{rgb}{0.5,0.5,0.5}
\definecolor{mauve}{rgb}{0.58,0,0.82}
\definecolor{lightblue}{rgb}{0.93,0.95,1.0}

\title{MindCross: Fast New Subject Adaptation with Limited Data for \\ Cross-subject Video Reconstruction from Brain Signals}
\author{
    Xuan-Hao Liu, Yan-Kai Liu, Tianyi Zhou, Bao-Liang Lu, Wei-Long Zheng\thanks{Corresponding author.}
}
\affiliations{
    School of Computer Science, Shanghai Jiao Tong University\\
    \{haogram\_sjtu, liu-yankai, tianyizhou, bllu, weilong\}@sjtu.edu.cn
}

\usepackage{bibentry}

\begin{document}

\maketitle

\begin{abstract}
Reconstructing video from brain signals is an important brain decoding task.
Existing brain decoding frameworks are primarily built on a subject-dependent paradigm, which requires large amounts of brain data for each subject. However, the expensive cost of collecting brain-video data causes severe data scarcity. Although some cross-subject methods being introduced, they often overfocus with subject-invariant information while neglecting subject-specific information, resulting in slow fine-tune-based adaptation strategy. To achieve fast and data-efficient new subject adaptation, we propose \textbf{MindCross}, a novel cross-subject framework. MindCross's \textit{N} specific encoders and one shared encoder are designed to extract subject-specific and subject-invariant information, respectively.
Additionally, a Top-\textit{K} collaboration module is adopted to enhance new subject decoding with the knowledge learned from previous subjects' encoders. Extensive experiments on fMRI/EEG-to-video benchmarks demonstrate MindCross's efficacy and efficiency of cross-subject decoding and new subject adaptation using only one model.
\end{abstract}

\begin{links}
    \link{Code}{https://github.com/XuanhaoLiu/MindCross}
\end{links}

\section{Introduction}
Our daily visual perceptions are composed of a series of seamless scenes, it is of great importance to investigate the neurologic mechanism of how human brain perceives and processes such dynamic visual perceptions \cite{makeig2002dynamic}.
One of brain decoding tasks is reconstructing the video from brain signals, which garnered significant interest recently \cite{yu2025mindpainter, sun2025neuralflix, gong2025mindtuner}.
These research deepens our understanding of brain function and offers promising advances for brain-computer interfaces (BCIs).

Spurred by the rapid development of deep multimodal models such as CLIP \cite{radford2021learning} and Stable Diffusion \cite{rombach2022high}, numerous previous works demonstrate the ability to reconstruct high-fidelity visual perception from brain signals \cite{benchetrit2024brain, li2024visual, scotti2023reconstructing} such as functional magnetic resonance imaging (fMRI) and electroencephalogram (EEG), which indirectly measures neural activity by detecting changes in blood oxygenation and spontaneous electrical activity from the scalp.

\begin{figure}[t]
    \centering
    \includegraphics[width=1.0\linewidth]{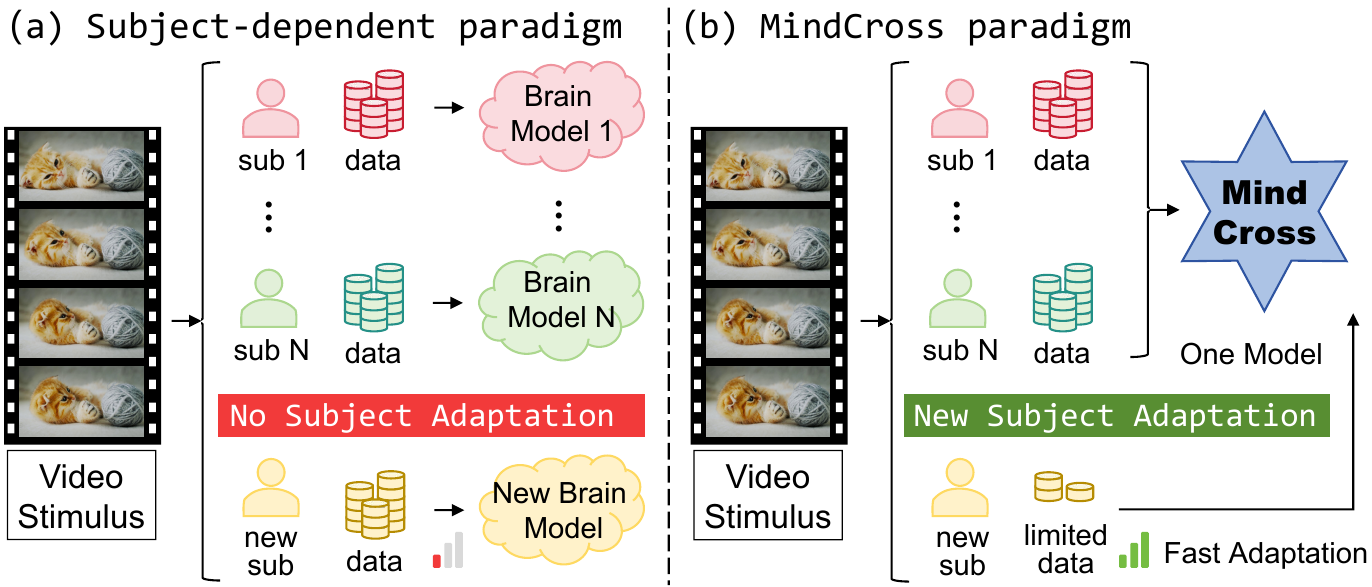}
    \caption{Illustration of cross-subject brain decoding: (a) Subject-dependent paradigm: training a particular model for each subject. (b) MindCross paradigm: training a universal model for all subjects, which can fastly adapt to new subjects with limited data.}
    \label{fig:illus}
\end{figure}

Unfortunately, there is a significant limitation in nowadays brain decoding frameworks that inherently restricts their wide applications.
As depicted in Figure \ref{fig:illus} (a), current works predominantly adhere to the subject-dependent paradigm \cite{chen2024bridging, chen2024mind, chen2025mindgpt}.
This paradigm necessitates extensive per-subject data acquisition and prolonged model training durations, thereby posing great constraints on its adaptability to new subjects and motivating us to develop more efficient alternatives.
Specifically, the cross-subject paradigm emerges as a promising solution, showing two key advantages: (1) decode neural patterns across multiple subjects using a single unified model, and (2) rapid adaptation for new subjects.

However, there are many difficulties in cross-subject brain decoding: \textbf{1) Subject Variability} Different subjects have different brain responses to the same stimuli, especially when processing high-level semantic information which requires complex cognition progress. \textbf{2) Data Scarcity for New Subject} The high expense of time and resources to conduct BCI experiments limits the amount of new subject data. \textbf{3) Utilization of Existing Subjects Data} Since existing subjects' brain data are always larger and more diverse than that of the new subject, it is a waste not to adequately utilize their valuable data.

Several cross-subject frameworks have been designed on an image brain decoding dataset \cite{allen2022massive}, whose subjects possess 27750 trials. However, since videos are viewed longer than images, the brain-video benchmark exhibits more significant data scarcity. For instance, each subject of the EEG-video benchmark only has 1400 trials.
Furthermore, previous cross-subject methods mostly adopt a fine-tune-based adaptation strategy \cite{scotti2024mindeye2, wang2024mindbridge}, which costs a lot of time.
These above issues motivate us to design a fast and data-efficient framework.

In this paper, we devise ``\textbf{MindCross}",  a novel cross-subject brain decoding framework designed for rapid adaptation to new subjects. MindCross addresses key challenges through three core innovations:
\textbf{1) Shared-specific Encoder Architecture} MindCross learns subject-invariant information by a shared encoder and allocates each subject a specific encoder to learn subject-related information. Through this architecture, MindCross is able to decouple learning subject-invariant and subject-related information to solve subject variability. \textbf{2) Fast New Subject Calibration} MindCross can rapidly adapt to him/her by only updating the parameter of the new subject while other modules are all frozen. This calibration method provides an efficient and effective solution for new subject adaptation while not harming the decoding performance of the existing subjects, solving data scarcity. \textbf{3) Collaboration Decoding} When the new subject has limited training data, MindCross decodes the semantic embeddings not only from the specific encoder of the new subject, but also from the encoders of Top K existing subjects similar to the new subject. This Top-K Collaboration Module largely helps MindCross to decode more precisely by adequately utilizing existing subjects' data.

In conclusion, our contributions are as follows:
\begin{itemize}
    \item We design MindCross, a shared-specific encoder architecture for decoupling learning subject-invariant and subject-related information.
    \item We design a novel calibration method for new subject adaptation that only needs updating a small number of parameters for the new subject while keeping other parts frozen, thus is fast and will not affect the decoding performance of existing subjects.
    \item We design a novel collaboration decoding module that utilizes existing subjects' data to help decoding a new subject with limited training data.
    \item Extensive experiments on fMRI/EEG-to-video benchmarks demonstrate the efficacy of MindCross.
\end{itemize}

\section{Related Work}

\subsection{Brain Decoding}
Researchers have been trying to decode visual perception (e.g., shape, color, and position) from brain activities for decades \cite{miyawaki2008visual}.
The rapid development of large generation models \cite{rombach2022high} enables high-fidelity images/videos generation, bringing a game-changing technique into the brain decoding field.
The current brain decoding pipeline \cite{chen2023cinematic, liu2024eegvideo, liu2025eegmirror, sun2025neuralflix, lu2025animate} can be summarized in two steps: mapping brain signals to CLIP embeddings and then using these embeddings to guide generation models in generating reconstructed images/videos.
However, most works fall into the subject-dependent fashion, leaving cross-subject brain decoding largely unexplored.

\subsection{Cross-subject Algorithm}
Cross-subject brain decoding has always been a hot topic because of the practical requirement of quickly adapting a BCI model to new subjects \cite{liu2024moge, liu2025mixeeg}.
For visual decoding task, MindBridge assigns each subject a particular encoder and adopts a cyclic auto-encoder training strategy to align different subject's brain features \cite{wang2024mindbridge}.
GLFA \cite{li2024enhancing}, STTM \cite{liu2025see} and MindAligner \cite{dai2025mindaligner} applies the functional alignment to align different subject's brain features.
MindEye2 \cite{scotti2024mindeye2} and MindTuner \cite{gong2025mindtuner} implicitly aligned all subject brain signals to a shared latent space.
Wills Aligner used a mixture of brain experts adapter to achieve few-shot learning \cite{bao2025wills}.
These methods put excessive attention on the subject-invariant information and fail to capture subject-related information, as their loss function encourages the model to learn similar brain representations from different subjects.

Moreover, for new subject adaptation, GLFA and MindTuner directly fine-tunes the whole model, while MindBridge freezes the deep layers and fine-tunes the shallow layers.
Such fine-tune-based strategies not only harm the decoding performance of the previous subjects but also cost a long time as they all update the encoders of previous subjects.
In contrast, MindCross achieves fast and data-efficient new subject adaptation with a novel calibration phase, where we update the new subject's specific encoder.
For the following comparison, we select the most representative cross-subject framework from image (MindBridge) and video (GLFA) filed.


\begin{figure*}[t]
    \centering
    \includegraphics[width=0.99\textwidth]{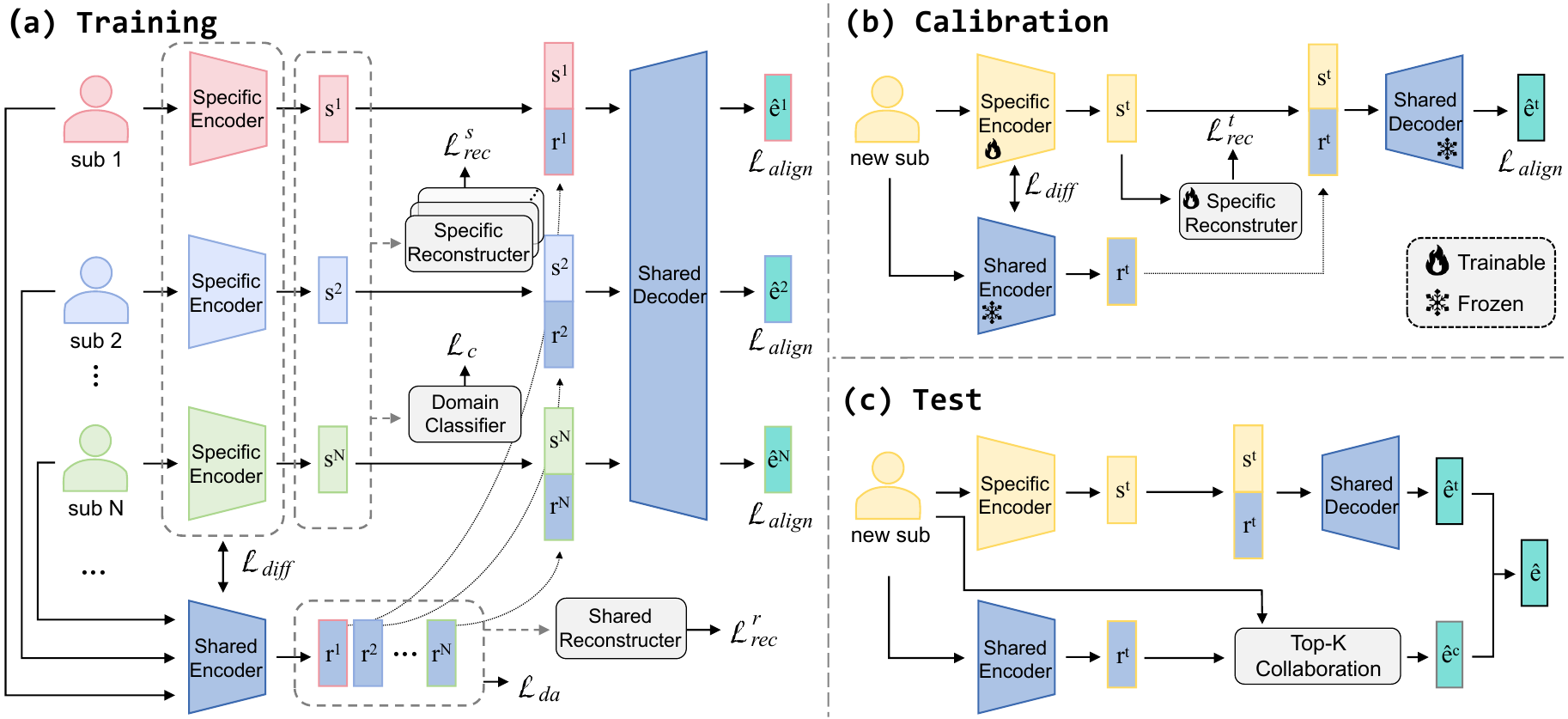}
    \caption{The framework of proposed MindCross consisting of  training, calibration, and test phase. \textbf{(a)} In the training phase, each specific encoder and shared encoder are optimized by several loss functions. \textbf{(b)} In the calibration phase, only the specific encoder of the new subject, marked with a flame icon, will be updated. \textbf{(c)} In the test phase, the final predictions are obtained from shared decoder and Top-\textit{K} Collaborate module.}
    \label{fig:method}
\end{figure*}

\subsection{Video Generation Models}
With large-scale image-text pair datasets \cite{schuhmann2022laionb}, nowadays diffusion models have demonstrated superior performance in the task of text-to-image generation \cite{rombach2022high}.
Compared to text-to-image (T2I) models \cite{mirza2014conditional}, text-to-video (T2V) generation models must maintain the temporal consistency between each frame.
Tune-A-Video \cite{wu2023tune} design an inflated network technique by adding a temporal attention layer in Unet to augment a T2I model to a T2V model, which was used as the video generation module in many previous brain decoding work \cite{chen2023cinematic, liu2024eegvideo, lu2025animate}. However, Tune-A-Video needs to fine-tune on the training set of video clips, consuming lots of time and computation resources.
Recently, more wonderful T2V models are proposed, such as AnimateDiff \cite{guo2024animatediff} and PyramidFlow \cite{jin2025pyramidal}.
In this paper, we focus on learning cross-subject brain representations, rather than taming SOTA T2V models for brain decoding, thus we simply used these off-the-shelf T2V models without further modification.

\section{Methodology}

\subsection{Overall Architecture}
Current methods of video generation from brain signals all follow a subject-dependent paradigm, requiring a large amount of data from one subject.
To overcome inter-subject variability and achieve rapid adaptation on new subjects, we propose MindCross, a shared-specific feature based framework inspired by ShaSpec \cite{wang2023multi}.
Depicted in Figure \ref{fig:method}, the whole framework consists of three phases: training, calibration, and test. 
In the training phase, \textit{N} specific encoders and one shared encoder are trained on each subject's data. 
If a new subject with limited data comes, then in the calibration phase, a new specific encoder will be trained while other modules in MindCross freezing. That is, we only update the new specific encoder and reconstructer.
In the test phase, the new subject's final output are obatined by its own specific encoder and a Top-\textit{K} Collaboration module.
Let us denote the \textit{N} subjects' data as $\textbf{X}=\{\textbf{x}_j^{i}, \textbf{y}_j^{i}\}_{i=1}^N$, where $\textbf{x}_j^{i}$ represents the $j^{th}$ data sample and the superscript $^i$ indexes subject. To simplify the notation, we omit the subscript $j$ when the contextual information is clear enough.
The specifc and shared encoder, reconstructer, and decoder are all MLP-like networks.


\subsection{Training Phase}
\subsubsection{Semantic Prediction}
As depicted in Figure \ref{fig:method} (a), \textit{N} specific encoders and one shared encoder are trained on each subject's data respectively.
The intuition behind the design of the specific and shared encoders is clear, we want the specific encoders to focus on the subject-related information and the shared encoders to focus on the subject-invariant information.
To achieve this, we introduce three losses here: the domain classification loss, domain alignment loss, and difference loss.
The process starts with the shared and specific branches running in parallel, with
\begin{equation}
    \textbf{r}^i = \textbf{E}_r(\textbf{x}^i),\  \textup{and}\ \  \textbf{s}^i = \textbf{E}_s(\textbf{x}^i),\ \ i \in \{1, \dots, N\}.
\end{equation}
Then the specific feature and the shared feature will be fused and fed into the universal decoder to calculate the semantic predictions (text CLIP embeddings):
\begin{equation}
    \hat{\textbf{e}}^i = \textbf{D}(\textup{ResFuse}(\textbf{s}^i, \textbf{r}^i)),
\end{equation}
where ResFuse stands for the residual fusion module: the specific and shared feature are first concatenated as the input of a projection layer $f$, whose output is added as a residual to the shared features to form the semantically rich brain embedding, as follows: 
\begin{equation}
    \textup{ResFuse}(\textbf{s}^i, \textbf{r}^i) = f(\textup{concat}(\textbf{s}^i, \textbf{r}^i)) + \textbf{r}^i.
\end{equation}

\subsubsection{Reconstruction Loss}
To force each encoder to extract the features that contain sufficient information from the original brain data, we introduce an auto-encoder architecture. That is, there are some reconstructers to reconstruct the original brain data from extracted features:
\begin{equation}
    \hat{\textbf{x}}_r^i = \textbf{R}_r(\textbf{r}^i),\  \textup{and}\ \  \hat{\textbf{x}}_s^i = \textbf{R}_s(\textbf{s}^i),
\end{equation}
we adopt the mean squared error (MSE) to calculate the reconstruction loss $\mathcal{L}_{rec}$:
\begin{equation}
    \mathcal{L}_{rec} = \mathcal{L}^s_{rec} + \mathcal{L}^r_{rec} = \frac{1}{N}\sum_i^N (\frac{1}{m} \Vert \hat{\textbf{x}}_s^i - x^i\Vert _2^2 + \frac{1}{m} \Vert \hat{\textbf{x}}_r^i - x^i\Vert _2^2),
\end{equation}
where \textit{m} is the length of the original brain data and $\Vert \Vert_2$ is the squared $L_2$-norm.

\subsubsection{Domain Classification Loss}
A domain classifier $\textbf{C}_{dc}$ is applied to predict which subject the specific feature was extracted from: $\hat{\textbf{y}}_{dc} = \textbf{C}_{dc}(\textbf{s}^i)$,
the domain classification loss is calculated using cross-entropy loss:
\begin{equation}
    \mathcal{L}_c = -\sum_{i=1}^N\sum_{l=1}^N \textbf{y}^i_{dc}\log \hat{\textbf{y}}^i_{dc},
\end{equation}
where $\textbf{y}_{dc}$ is the domain label, i.e., an one-hot vector indicate the subject index.

\subsubsection{Domain Alignment Loss}
\label{sec:dal}
In contrast to the specific encoder, the shared encoder is expected to extract the subject-invariant information. Hence, domain labels are supposed to be removed from shared features $\textbf{r}^i, \ i\in\{1, \dots, N\}$. That is, in an ideal situation, it is impossible to distinguish which subject the shared feature was extracted from.
One way to achieve this is to confuse the domain classifier by minimizing the cross-entropy loss via a gradient reversal layer (GRL) \cite{ganin2015unsupervised}:
\begin{equation}
    \mathcal{L}_{da} = \sum_{i=1}^N\sum_{l=1}^N \textbf{y}^i_{da}\log \hat{\textbf{y}}^i_{da},
\end{equation}
where $\textbf{y}_{da}$ is the domain label, and $\hat{\textbf{y}}_{da} = \textbf{C}_{da}(\textbf{r}^i)$, $\textbf{C}_{da}$ is another domain classifier for domain alignment. The GRL acts as identity transform for forwardprop, and reverses the gradient for backprop.



\subsubsection{Difference Loss}
Through the shared-specific design, a subject's brain data is processed by two encoders.
There will be a waste of computation if both of the specific and the shared encoders extracted very similar information from one brain data.
Thus, we adopt the Hadamard product $\odot$ to enhance the orthogonality and simplify the difference loss as follows:
\begin{equation}
     \mathcal{L}_{diff} = \frac{1}{N}\sum_i^N \Vert \textbf{s}^i \odot \textbf{r}^i \Vert_2^2.
\end{equation}

\subsubsection{Alignment Loss}
MindCross adopts the contrastive loss and MSE loss to align the semantic predictions with text CLIP embeddings.
In order to focus the research on cross-subject brain decoding, we only discuss the simplified situation of only predicting the text CLIP embedding rather than both frame latents and text embeddings.
Although previous works have shown that the predicted frame latents help to improve low-level metrics \cite{liu2024eegvideo, gong2024neuroclips, lu2025animate}, overall is not critical in our study and we omit this part for a more focused paper (since this part is just adding an additional decoder to predict frame latents from the same feature used to predict the text CLIP embeddings).

For contrastive loss, MindCross adopt the SoftCLIP loss which was introducted in MindEye \cite{scotti2023reconstructing}:
\begin{equation}
\centering
\begin{split}
\mathcal{L}_{\mathrm{SoftCLIP}}(\textbf{e}, \hat{\textbf{e}}) &= -{\sum\nolimits_{k=1}^B} {\sum\nolimits_{l=1}^B} \\
[{\frac{\mathrm{exp}(\textbf{e}_{k}\cdot \textbf{e}_{l}) / \tau}
{\sum\nolimits_{m=1}^B \mathrm{exp}({\textbf{e}_{k}\cdot \textbf{e}_{m}}/{\tau})}}
&\cdot \log ({\frac{\mathrm{exp}(\hat{\textbf{e}}_{k} \cdot \textbf{e}_{l}) / \tau}
{\sum\nolimits_{m=1}^B \mathrm{exp}({\hat{\textbf{e}}_{k}\cdot \textbf{e}_{m}}/{\tau})}})],
\end{split}
\end{equation}
where \textbf{e} and $\hat{\textbf{e}}$ are the ground truth text embedding and the semantic predictions in a batch of size \textit{B}. $\tau$ is a temperature hyperparameter.
Besides the contrastive loss for better clustering of different classes, MindCross adds an MSE loss to improve alignment and guarantee the generative quality of the video:
\begin{equation}
    \mathcal{L}_{align} = \mathcal{L}_{\mathrm{SoftCLIP}}(\textbf{e}, \hat{\textbf{e}}) + \frac{1}{N}\sum_i^N \Vert \textbf{e}^i- \hat{\textbf{e}}^i \Vert_2^2.
\end{equation}

Finally, MindCross is trained end-to-end by incorporating all
these losses to achieve cross-subject brain decoding. These Greek letters are all hyperparameters for balancing each loss function:
\begin{equation}
    \mathcal{L}_{train} = \mathcal{L}_{align} + \alpha \mathcal{L}_{rec} + \beta \mathcal{L}_{c} + \gamma \mathcal{L}_{da} + \zeta\mathcal{L}_{diff}.
\end{equation}

\begin{figure}[t]
    \begin{center}
        \includegraphics[width=\linewidth]{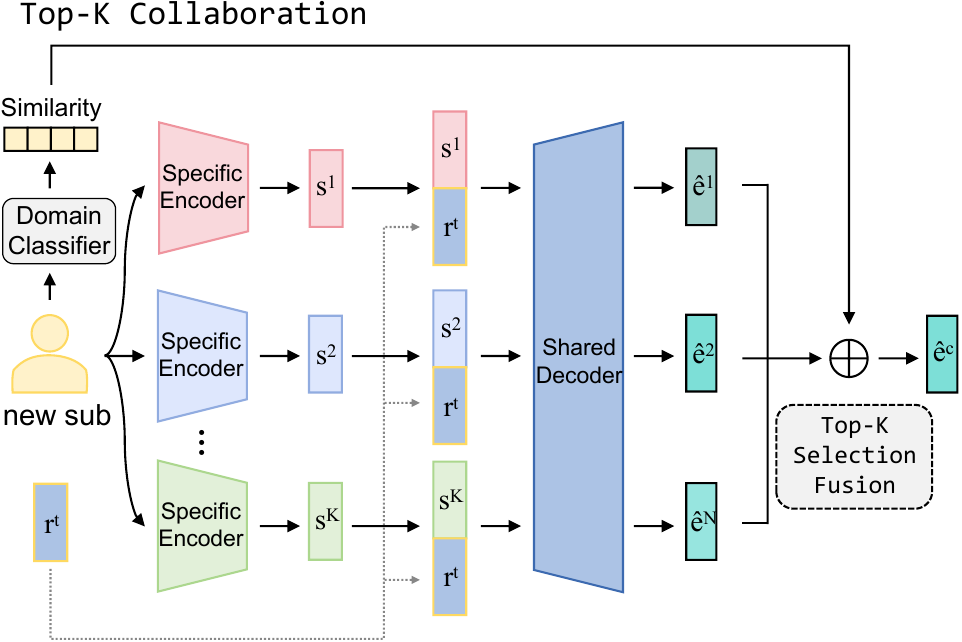}
        \caption{Top-\textit{K} Collaborate Module: The similarity vector is obtained by feeding the new subject's feature $s^t$ into the domain classifier, then we select the Top-\textit{K} similar domains for collaborating to calculate the final output.}
        \label{fig:topk}
    \end{center}
\end{figure}

\subsection{Calibration Phase}
If there comes a new subject, the traditional per-subject per-model paradigm needs to train a new brain decoding model from scratch, which is time and computation consuming and requires massive data from the new subject.
Our MindCross propose a calibration way to fully utilize the universal model trained on previous subjects' brain data to achieve fast new subject adaptation.
Specifically, we freeze the previously trained MindCross and only update the specific encoder and reconstructer of the new subject.
By freezing most of the parameters, we save lots of time and computation resources.
Most importantly, it is possible to achieve comparative decoding results with limited data, as the shared encoder and decoder are pre-trained on a large amount of subjects' brain data.
The calibration loss is written as follows, each loss function is defined in the previous section:
\begin{equation}
    \mathcal{L}_{calib} = \mathcal{L}_{align} + \alpha' \mathcal{L}^{t}_{rec} + \beta'\mathcal{L}_{diff}.
\end{equation}

\subsection{Test Phase}
In the test phase, MindCross predict the text CLIP embeddings by two branches: one uses the specific encoder of the new subject $\textbf{E}_t$, and another one utilizes the trained specific encoders $\{\textbf{E}^{i}_{s}\}_{i=1}^N$ through a \textbf{Top-\textit{K} Collaboration Module} as depicted in Figure \ref{fig:topk}.
Firstly, the specific feature of the new subject $\textbf{s}^t$ is fed into the domain classifier $\textbf{C}_{dc}$ to calculate the similarity $p$ between the brain data of new subject and previous subjects. We select the Top-\textit{K} subjects' specific encoders and predict the text CLIP embeddings. Afterwards, the \textit{K} predictions are added together according to the similarity weights. The higher weight indicates that the distribution is more similar to the new subject, so more trust can be given to the corresponding prediction.
\begin{equation}
    \hat{\textbf{e}}^c = \sum_{k\in\mathrm{TopK}(p)} p_k \cdot \hat{\textbf{e}}^k,
\end{equation}
where $\mathrm{TopK}$ selects the top \textit{K} index in a logits vector $p$. Finally, the semantic prediction of the new subject is combining $\hat{\textbf{e}}^c$ and $\hat{\textbf{e}}^t$ as demonstrated in Figure \ref{fig:method}, $\lambda=1e-2$:
\begin{equation}
    \hat{\textbf{e}} = \hat{\textbf{e}}^t + \lambda \hat{\textbf{e}}^c.
\end{equation}

\subsection{Video Generation Module}
The video generation model plays a crucial roles in all brain decoding works as pre-training text-to-video (T2V) diffusion models possess a large amount of prior knowledge from the graphics, image, and video domains.
However, as demonstrated before, our research focus is not on taming the cut-the-edge T2V models for generating smooth and high-fidelity videos, but on how to learn cross-subject brain representations. Therefore, we adopted a T2V models called PyramidFlow \cite{jin2025pyramidal} for video generation without further modification. Specifically, the videos are generated using the semantic prediction $\hat{\textbf{e}}$ as the condition.

\section{Experiments}

\begin{table*}[t]
\setlength\tabcolsep{3.3pt}
\centering
\small
\begin{tabular}{l|lc|cccccccc}
    \toprule
    ~ & \multirow{3}[3]{*}{Method} & \multirow{3}[3]{*}{Venue} & \multirow{3}[3]{*}{\# Models} & \multicolumn{3}{c}{Video-based} & \multicolumn{4}{c}{Frame-based} \\
    \cmidrule(lr){5-7} \cmidrule(lr){8-11}
    ~ & & & &  \multicolumn{2}{c}{Semantic-Level} & \multicolumn{1}{c}{ST-Level} & \multicolumn{2}{c}{Semantic-Level} & \multicolumn{2}{c}{Pixel-Level}\\
    \cmidrule(lr){5-6} \cmidrule(lr){7-7} \cmidrule(lr){8-9} \cmidrule(lr){10-11} 
    ~ & & & &  2-way    & 40/50-way  & CLIP-pcc  & 2-way   & 40/50-way & SSIM   & PSNR    \\
    \midrule
    \multirow{7}*{\rotatebox{90}{SEED-DV}} & MinD-Video & NeurIPS 23   & 20        &    0.805\scriptsize{$\pm0.02$}        &        0.156\scriptsize{$\pm0.03$}      &    0.411\scriptsize{$\pm0.47$}      &       0.755\scriptsize{$\pm0.03$}     &   0.128\scriptsize{$\pm0.03$}     &     0.176\scriptsize{$\pm0.06$}     &       8.579\scriptsize{$\pm1.42$}  \\
    ~ & NeuroClips & NeurIPS 24     &    20   &    \underline{0.809\scriptsize{$\pm0.03$}}        &        0.154\scriptsize{$\pm0.03$}      &    \underline{0.756\scriptsize{$\pm0.28$}}      &       \underline{0.785\scriptsize{$\pm0.03$}}     &   0.151\scriptsize{$\pm0.04$}     &     0.238\scriptsize{$\pm0.08$}     &       \underline{8.703\scriptsize{$\pm1.37$}}  \\
    ~ & Mind-Animator & ICLR 25 & 20         &    0.799\scriptsize{$\pm0.03$}        &        0.158\scriptsize{$\pm0.02$}      &    0.421\scriptsize{$\pm0.56$}      &       0.768\scriptsize{$\pm0.02$}     &   0.142\scriptsize{$\pm0.04$}     &     0.253\scriptsize{$\pm0.12$}     &       8.679\scriptsize{$\pm1.52$}  \\
    ~ & EEG2Video & NeurIPS 24 & 20       &    0.800\scriptsize{$\pm0.03$}        &        \underline{0.161\scriptsize{$\pm0.01$}}      &    0.412\scriptsize{$\pm0.45$}      &       0.772\scriptsize{$\pm0.03$}     &   \underline{0.146\scriptsize{$\pm0.01$}}     &     \underline{0.258\scriptsize{$\pm0.08$}}     &       8.684\scriptsize{$\pm1.46$}  \\
    \cmidrule(lr){2-11}
    ~ & GLFA  & ECCV 24   & 1         &    0.778\scriptsize{$\pm0.02$}        &        0.152\scriptsize{$\pm0.02$}      &    0.751\scriptsize{$\pm0.52$}      &       0.743\scriptsize{$\pm0.03$}     &   \textbf{0.136\scriptsize{$\pm0.03$}}     &     0.192\scriptsize{$\pm0.08$}     &       8.642\scriptsize{$\pm1.48$}  \\
        ~ & MindBridge  & CVPR 24               & 1           &    0.782\scriptsize{$\pm0.03$}        &        0.148\scriptsize{$\pm0.02$}      &    0.753\scriptsize{$\pm0.43$}      &       0.749\scriptsize{$\pm0.03$}     &   0.125\scriptsize{$\pm0.02$}     &     0.185\scriptsize{$\pm0.07$}     &       8.625\scriptsize{$\pm1.39$}  \\
    \rowcolor{lightblue}
    ~ & MindCross & Ours             & 1     &    \textbf{0.786\scriptsize{$\pm0.02$}} &  \textbf{0.154\scriptsize{$\pm0.03$}}  &    \textbf{0.758\scriptsize{$\pm0.32$}}   &       \textbf{0.752\scriptsize{$\pm0.03$}}     &   0.128\scriptsize{$\pm0.02$}     &     \textbf{0.197\scriptsize{$\pm0.06$}}     &       \textbf{8.658\scriptsize{$\pm1.43$}}  \\
    \midrule
    \multirow{7}*{\rotatebox{90}{CC2017}} & MinD-Video & NeurIPS 23   & 3         & \underline{0.839\scriptsize{$\pm0.03$}} & {0.197\scriptsize{$\pm0.02$}} & {0.408\scriptsize{$\pm0.46$}}    & {0.796\scriptsize{$\pm0.03$}} & {0.174\scriptsize{$\pm0.03$}}  & {0.171\scriptsize{$\pm0.08$}} & 8.662\scriptsize{$\pm1.52$}      \\ 
    ~ & NeuroClips & NeurIPS 24     &     3&    0.834\scriptsize{$\pm0.03$}        &        \underline{0.220\scriptsize{$\pm0.01$}}      &    \underline{0.738\scriptsize{$\pm0.17$}}      &       0.806\scriptsize{$\pm0.03$}     &   \underline{0.203\scriptsize{$\pm0.01$}}     &     \underline{0.390\scriptsize{$\pm0.08$}}     &       9.211\scriptsize{$\pm1.46$}  \\
    ~ & Mind-Animator  & ICLR 25 & 3         &         0.830\scriptsize{$\pm0.02$}          &       0.185\scriptsize{$\pm0.04$}         &     0.425\scriptsize{$\pm0.52$}          &             \underline{0.816\scriptsize{$\pm0.03$}}      &      0.182\scriptsize{$\pm0.03$}        &        0.321\scriptsize{$\pm0.12$}         &          \underline{9.220\scriptsize{$\pm1.48$} }             \\
    ~ & EEG2Video & NeurIPS 24 & 3     & {0.833\scriptsize{$\pm0.03$}} & {0.209\scriptsize{$\pm0.02$}} & {0.413\scriptsize{$\pm0.37$}}    & {0.811\scriptsize{$\pm0.04$}} & {0.191\scriptsize{$\pm0.03$}}  & {0.318\scriptsize{$\pm0.14$}} & 8.763\scriptsize{$\pm1.45$}      \\
    \cmidrule(lr){2-11}
    ~ & GLFA  & ECCV 24   & 1       & \textbf{0.840\scriptsize{$\pm0.03$}} & {0.209\scriptsize{$\pm0.02$}} & {0.742\scriptsize{$\pm0.49$}}    & {0.805\scriptsize{$\pm0.04$}} & {0.187\scriptsize{$\pm0.03$}}  & {0.134\scriptsize{$\pm0.07$}} & 8.763\scriptsize{$\pm1.45$}      \\ 
        ~ & MindBridge  & CVPR 24               & 1         & {0.821\scriptsize{$\pm0.03$}} & {0.206\scriptsize{$\pm0.02$}} & {0.754\scriptsize{$\pm0.42$}}  & \textbf{0.813\scriptsize{$\pm0.03$}} & {0.195\scriptsize{$\pm0.02$}}  & {0.156\scriptsize{$\pm0.07$}} & 8.972\scriptsize{$\pm1.67$}      \\ 
    \rowcolor{lightblue}
    ~ & MindCross & Ours             & 1   & 0.830\scriptsize{$\pm0.03$} & \textbf{0.210\scriptsize{$\pm0.03$}} & \textbf{0.762\scriptsize{$\pm0.34$}} & {0.802\scriptsize{$\pm0.03$}} & \textbf{0.198\scriptsize{$\pm0.03$}}  & \textbf{0.163\scriptsize{$\pm0.07$}} & \textbf{8.983\scriptsize{$\pm1.54$}}      \\ 
    \bottomrule
\end{tabular}
\caption{\textbf{Quantitative comparison of brain decoding between MindCross and other methods.} The best performance of cross-subject frameworks are marked in BOLD. The best performance of per-subject per-model methods are underlined.}
\label{tab:cross}
\end{table*}

\subsection{Datasets}
In this study, we utilize two publicly available brain-video datasets, which includes paired stimulus videos and their corresponding brain responses. More details of each dataset are written in appendix.

For EEG-to-video reconstruction, we use the SEED-DV dataset \cite{liu2024eegvideo}. This dataset contains 20 subjects' EEG signals while they were watching 7 video blocks containing 1400 two-second video clips of 40 concepts, covering various animals, scenes, and activities.
We use the first 6 blocks (1200 trials) as the training set and the last block (200 EEG-video trials) as the test set.

For fMRI-to-video reconstruction, we use the CC2017 dataset \cite{wen2018neural} recording 3 subjects' fMRI data using a 3-T MRI system with two-second temporal resolution.
The dataset consists of a training set containing 18 8-minute video clips and a test set containing five 8-minute video clips. Each subject viewed the training and testing video clips 2 and 10 times, respectively, and the test set was averaged across trials. We divide them into two-second small clips. Thus there are 8640 training samples and 1200 testing samples of fMRI-video pairs.

\begin{figure}[t]
    \centering
    \includegraphics[width=1.0\linewidth]{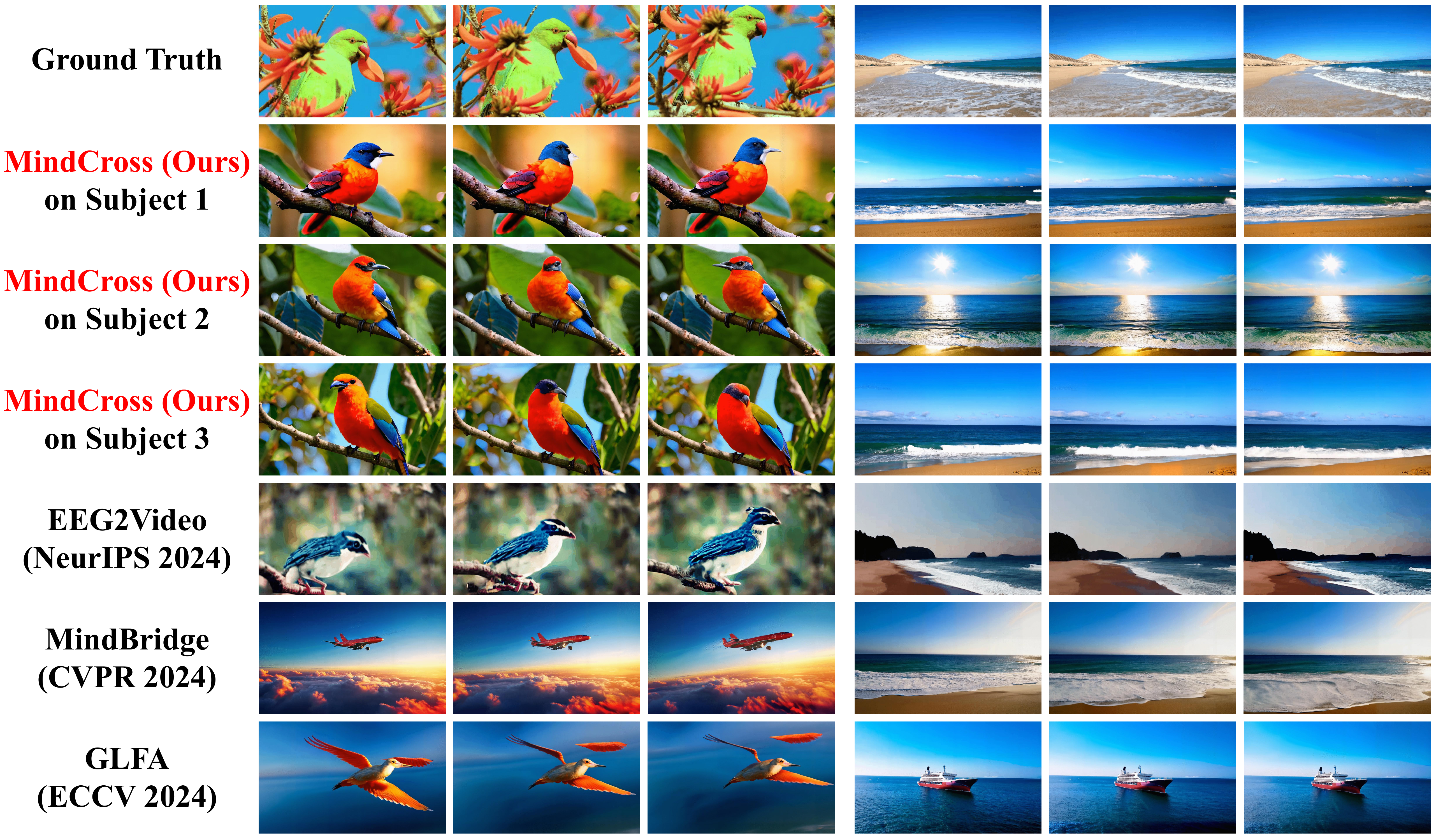}
    \caption{Comparison on EEG-to-video benchmark.}
    \label{fig:qua_seeddv}
\end{figure}

\begin{figure}[t]
    \centering
    \includegraphics[width=1.0\linewidth]{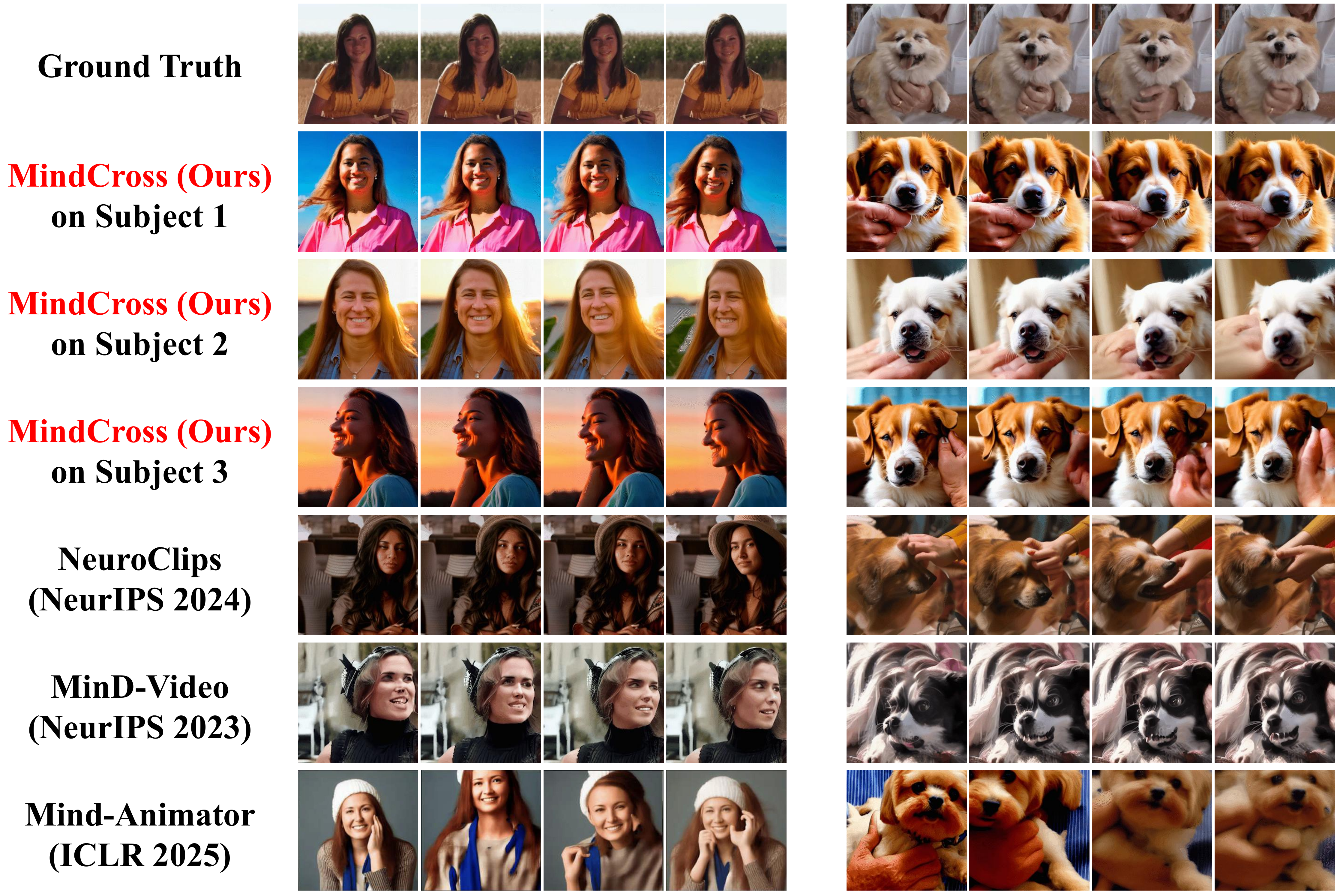}
    \caption{Comparison on fMRI-to-video benchmark.}
    \label{fig:qua_cc2017}
\end{figure}




\subsection{Evaluation Metrics}
Following previous work \cite{gong2024neuroclips}, we perform quantitative evaluation using both frame-based and video-based metrics. Frame-based metrics evaluate each frame individually, providing a snapshot evaluation, while video-based metrics evaluate the quality of the video, emphasizing the consistency and smoothness.
For semantic evaluation, we use the $N$-way top-$K$ metric and set $K$ to 1, which means a video is considered successfully reconstructed if the ground truth (GT) class is in the top 1 probabilities using a pretrained classifier.
For frame-based metric, the classifier is a CLIP-based classifier \cite{radford2021learning} trained on ImageNet \cite{deng2009imagenet}. For video-based metric, the classifier is a VideoMAE-based \cite{tong2022videomae} video classifier trained on Kinetics-400 dataset \cite{kay2017kinetics}.
For spatiotemporal-level metrics to measure video consistency, we compute the average cosine similarity between all pairs of adjacent video frames' CLIP image embeddings, CLIP-pcc.
The structural similarity index metric (SSIM) and peak signal-to-noise ratio (PSNR) are used as pixel-level metrics.
More details of all metrics are written in appendix.

\subsection{Cross-Subject Video Reconstruction}
Our MindCross can achieve cross-subject brain decoding across several subjects with only one model, while most previous methods require to train a particular model for each subject. We compare its average video reconstruction performance across all subjects with that of state-of-the-art subject-dependent methods: MinD-Video \cite{chen2023cinematic}, NeuroClips \cite{gong2024neuroclips}, Mind-Animator \cite{lu2025animate}, EEG2Video \cite{liu2024eegvideo}, and two cross-subject methods: MindBridge \cite{wang2024mindbridge}, and GLFA \cite{li2024enhancing}. The quantitative and
qualitative results for all methods are presented in Table \ref{tab:cross}, and Figure \ref{fig:qua_seeddv}, and Figure \ref{fig:qua_cc2017} respectively. It can be seen that our MindCross achieve comparable performance against per-subject per-model methods while maintaining just one model in semantic level metrics, and outperforms other cross-subject methods, demonstrating our success on cross-subject brain decoding.
Figure \ref{fig:qua_seeddv} demonstrates our MindCross achieves more accurate in semantic decoding across different subjects, e.g., MindBridge decoded bird as plane, and GLFA decoded beach as ship.

\subsection{New-Subject Adaptation}
\begin{figure}[htbp]
    \centering
    \includegraphics[width=1.0\linewidth]{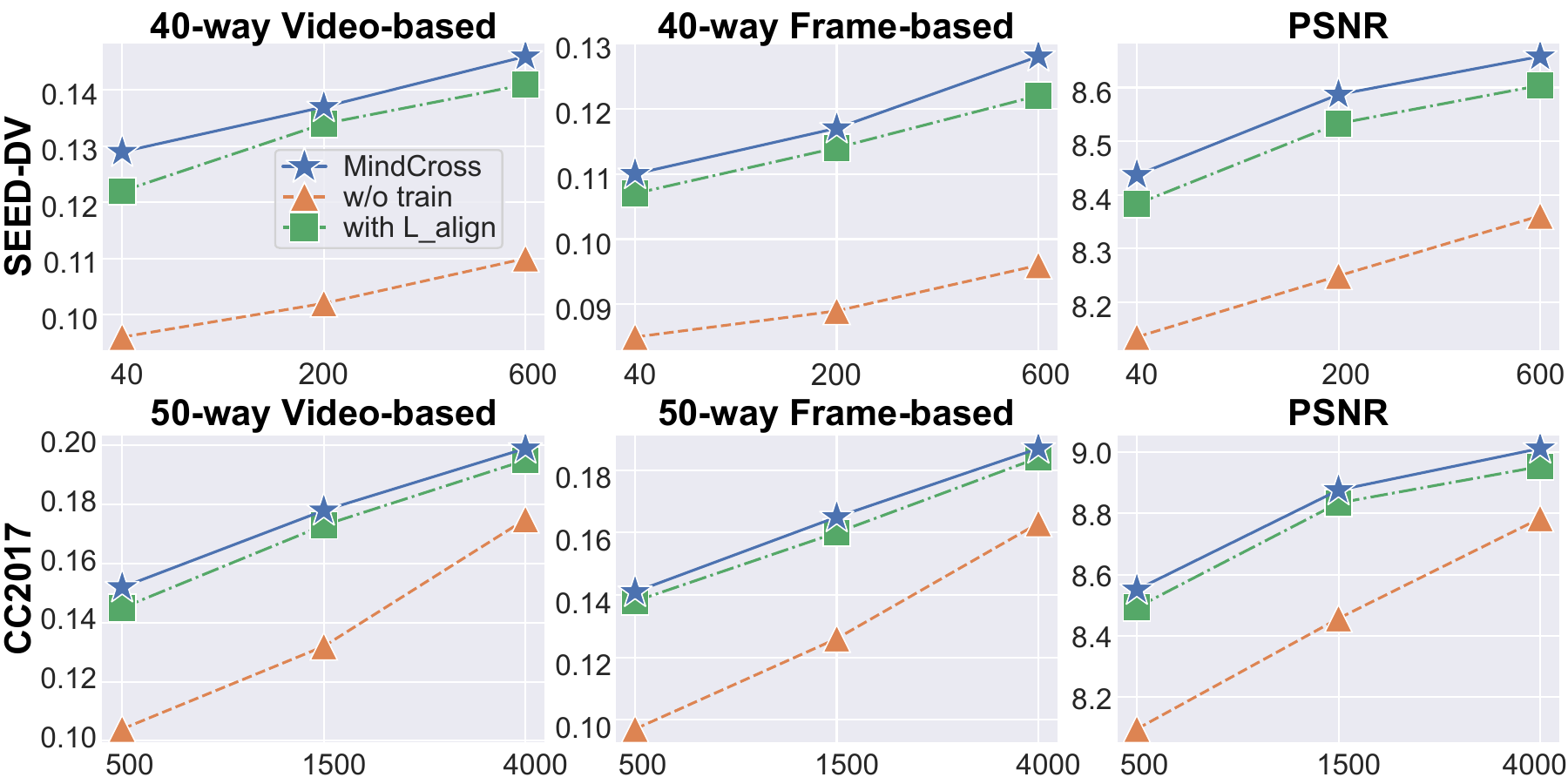}
    \caption{Results of new subject adaptation in limited data scenario. The x-axis is the number of data from new subjects, the y-axis is the metric.}
    \label{fig:newsub_line}
\end{figure}

MindCross can effectively transfer its pretrained knowledge to adapt to new subjects, offering significant advantages in real-world applications.
To simulate the case when the new subject only has limited data, we select 40, 200, and 600 samples from the new subjects for the EEG dataset and 500, 1500, and 4000 samples from the new subjects for the fMRI dataset.
We perform a leave-one-subject-out (LOSO) experiment here. The results are displayed in Figure \ref{fig:newsub_line}, where MindCross performs comparable on the new subject to the existing subjects.
MindCross significantly outperforms the variant `w/o train' (training MindCross on new subject from scratch), which indicates the efficacy of the training phase.
We also evaluate the performance of calibration with only $\mathcal{L}_{align}$. It can be observed that the difference loss and reconstruction loss improve the decoding results.
Furthermore, the comparison with baselines is displayed in Table \ref{tab:calibrate_time_size}, we calibrated on the new subject using 200/500 data on EEG/fMRI datasets, respectively. It can be seen that our MindCross significantly reduces the adapation time and updating parameter (especially when the number of subjects is large), while outperforming or tying the baselines, demonstrating the superiority of MindCross on new subject adaptation.

\begin{table}[H]
\setlength\tabcolsep{3.6pt}
\footnotesize
\centering
\begin{tabular}{l|lccccc}
    \toprule
    ~ & {Training Loss} & 40-V  & 40-F & PSNR & Time/sec & \# Para   \\
    \midrule
    \multirow{3}*{\footnotesize \rotatebox{90}{EEG}} & MindBridge	    &  \textbf{0.142}  &  0.104  &  8.514  &  \underline{5.104}  &  \underline{126.81M}\\
    ~ & GLFA & 0.135 & \textbf{0.121} & \underline{8.522}  &  10.651  &  247.27M\\
    ~ & MindCross & \underline{0.137}  &  \underline{0.117}  &  \textbf{8.587}  &  \textbf{1.090}  &  \textbf{9.77M} \\
    \midrule
    \multirow{3}*{\footnotesize \rotatebox{90}{fMRI}} & MindBridge   & 0.147  &  \textbf{0.145}  &  8.547  &  \underline{9.267}  &  \underline{293.43M}\\
    ~ & GLFA  & \textbf{0.153}  &  0.137  &  \textbf{8.601}  &  17.513  &  504.23M\\
    ~ & MindCross & \underline{0.152}  &  \underline{0.141}  &  \underline{8.551}  & \textbf{ 2.724}  &  \textbf{109.10M}\\
    \bottomrule
\end{tabular}
\caption{Comparison with other cross-subject baselines on new subject adaptation with limited data task.}
\label{tab:calibrate_time_size}
\end{table}

\subsection{Ablation Study}
\subsubsection{Ablation on Training Loss}
Table \ref{tab:ablation_train_loss} displays the results of introducing different losses in the training phase. It can be seen that MindCross with $\mathcal{L}_{da} + \mathcal{L}_{dc} + \mathcal{L}_{rec}$ outperforms MindCross with only $\mathcal{L}_{align}$, demonstrating the shared-specific architecture helps model learn cross-subject representations. Adding $\mathcal{L}_{diff}$ does not affect the decoding performance much, but is useful for new subject adaptation in the calibration phase. Thus, we add the difference loss in the training phase to encourage the shared-specific encoder focus on the different aspects of brain data.

\begin{table}[H]
\setlength\tabcolsep{4pt}
\footnotesize
\centering
\begin{tabular}{l|lcccc}
    \toprule
    ~ & {Training Loss} & 2-way-V  & 2-way-F  & SSIM  & PSNR    \\
    \midrule
    \multirow{3}*{\footnotesize \rotatebox{90}{EEG}} & $\mathcal{L}_{align}$     &    0.756        &        0.693     &    0.187   &    8.627\\
    ~ & $ + \mathcal{L}_{rec} + \mathcal{L}_{da} + \mathcal{L}_{dc}$   &    \textbf{0.789}        &        \textbf{0.757}      &    \underline{0.195}     &    \underline{8.647} \\
    ~ & $ + \mathcal{L}_{diff}$ (Ours)  &    \underline{0.786}        &        \underline{0.752}      &    \textbf{0.197}     &    \textbf{8.658} \\
    \midrule
    \multirow{3}*{\footnotesize \rotatebox{90}{fMRI}} & $\mathcal{L}_{align}$    &    0.812        &        0.785      &    0.152     &    8.576 \\
    ~ & $ + \mathcal{L}_{rec} + \mathcal{L}_{da} + \mathcal{L}_{dc}$  &    0.829        &        \textbf{0.810}      &    \textbf{0.168}     &    \underline{8.957} \\
    ~ & $ + \mathcal{L}_{diff}$ (Ours) &    \textbf{0.830}        &        \underline{0.802}      &    \underline{0.163}     &    \textbf{8.983} \\
    \bottomrule
\end{tabular}
\caption{Ablation of different losses at the training phase. Models are trained and tested across all subjects.}
\label{tab:ablation_train_loss}
\end{table}

\subsubsection{Ablation on Top-K Collaboration Module}
\begin{figure}[H]
    \centering
    \includegraphics[width=1.0\linewidth]{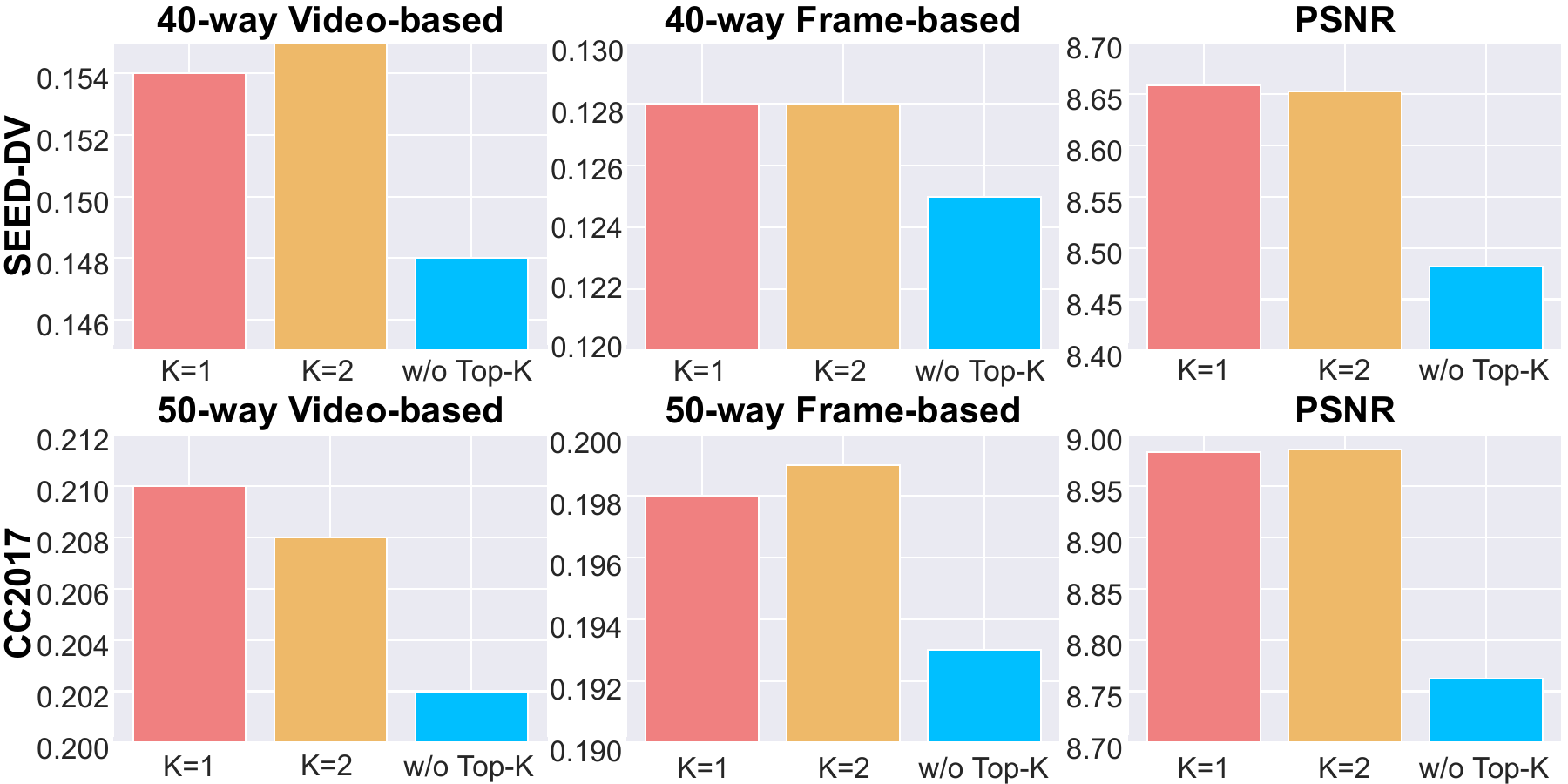}
    \caption{Ablation study of Top-K Collaboration Module.}
    \label{fig:topk_ab}
\end{figure}

In this case, we use the $\hat{\textbf{e}}^t$ to generate videos as the model `w/o Top-K'. The results are depicted in Figure \ref{fig:topk_ab}, which demonstrates the efficacy of the Top-K Collaboration Module. There is no significant difference between $K=1$ and $K=2$. Thus, we set $K=1$ in this paper.

\subsection{Visualization}
\subsubsection{Top-K Subject Selection}
\begin{figure}[htbp]
    \centering
    \includegraphics[width=0.67\linewidth]{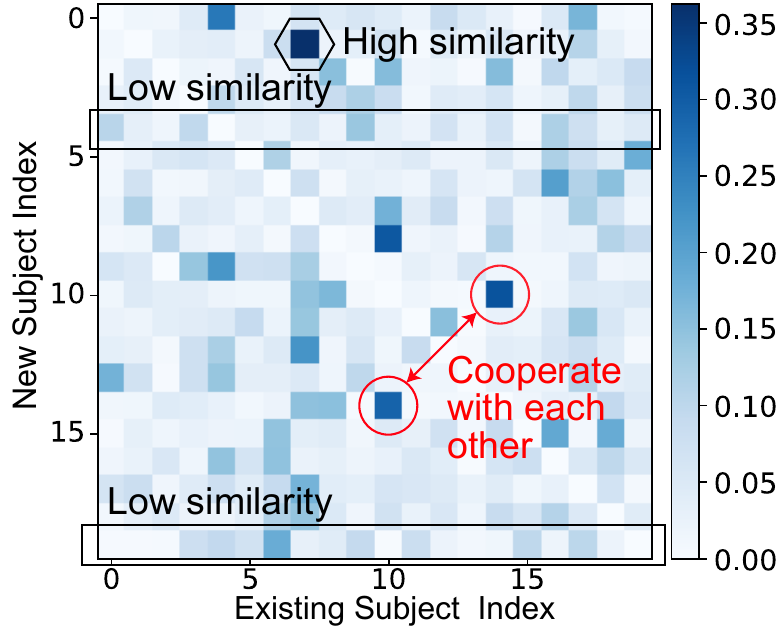}
    \caption{The heat map of the subject selection probability of the Top-K Collaboration Module.}
    \label{fig:topkvis}
\end{figure}

To demonstrate the effectiveness of the Top-K collaboration module, We enumerate the new subject in the SEED-DV dataset and train a MindCross model on the other 19 subjects. We visualize the subject selection probability of each subject in Figure \ref{fig:topkvis}, where $K=1$ in this case.
It can be seen that sub 10 and sub 14 will cooperate with each other with a high probability, which indicates it is useful to introduce a mechanism like memory retrieval process to recall the existing subjects' similar data.
Although some subjects like sub 4 and sub 19 have low similarity with all subjects, the Top-K module still improves the performance as shown in Figure \ref{fig:topk}.

\subsubsection{Shared and Specific Features}
\begin{figure}[H]
  \centering
    \includegraphics[width=0.9\linewidth]{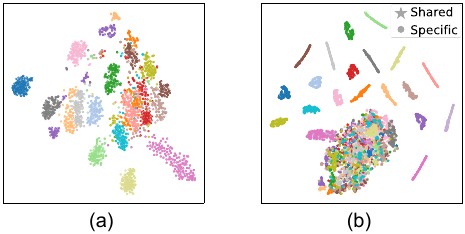}
  \caption{
  Feature visualization of MindCross using t-SNE. (a) Original brain data. The color represents different subjects. (b) Learned features of MindCross.
  }
  \label{fig:shaspecfeature}
\end{figure}

Our MindCross is supposed to separate the subject-related and subject-invariant components out from the original brain signals. Here we trained our model on 19 subjects from the SEED-DV dataset and obtained their original data $\textbf{x}^i$ along with the specific feature $\textbf{s}^i$ and shared feature $\textbf{r}^i$ learned by our MindCross.
The 19 subject's original brain data are displayed in Figure \ref{fig:shaspecfeature} (a), where their brain data differ significantly from each other. After the process of MindCross, the subject-related and subject-invariant components are successfully extracted, as shown in Figure \ref{fig:shaspecfeature} (b).
Compared to other previous cross-subject methods which only extract the subject-invariant information, our MindCross keeps the subject-related information. Therefore, we can calculate the similarity between new subject and existing subjects to enhance the brain decoding for new subject.

\section{Conclusion}
In this paper, we propose a cross-subject brain decoding framework \textbf{MindCross}, which can effectively extract both subject-invariant and subject-related information. Through a novel calibration phase and collaboration module, MindCross significantly reduces the adaptation time and updating parameter, demonstrating great superiority and efficiency in new subject adaptation with limited data.
Our work not only enhances cross-subject brain decoding, but also showcases promising practical applications in the BCI field.

\section{Acknowledgements}
This work was supported in part by grants from Brain Science and Brain-like Intelligence Technology-National Science and Technology Major Project (2025ZD0218900), National Natural Science Foundation of China (62376158), STI 2030-Major Projects+2022ZD0208500, Medical-Engineering Interdisciplinary Research Foundation of Shanghai Jiao Tong University “Jiao Tong Star” Program (YG2023ZD25, YG2024ZD25 and YG2024QNA03), Shanghai Jiao Tong University 2030 Initiative, the Lingang Laboratory (Grant No. LGL-1987), GuangCi Professorship Program of RuiJin Hospital Shanghai Jiao Tong University School of Medicine, and Shanghai Jiao Tong University SCS-Shanghai Emotionhelper Technology Co., Ltd Joint Laboratory of Affective Brain-Computer Interfaces.




\bibliography{aaai2026}

\clearpage
\appendix

\section{Dataset Details and Preprocessing}
\label{apd:dataset}
In this study, we evaluate our brain-to-video reconstruction experiments using the opensource fMRI-video dataset (cc2017 dataset) \cite{wen2018neural} and the EEG-video dataset (SEED-DV) \cite{liu2024eegvideo}. We segmented all video into 2-second clips, down-sampled the frame rate to 3 FPS (i.e. evenly select 6 frames).
Following the approach of MinD-Video \cite{chen2023cinematic}, we employed an image captioning model called BLIP2 \cite{li2022blip} to obtain the text prompt for each video clip using the first frame, with lengths not exceeding 20 words.

\subsection{EEG-to-video: SEED-DV dataset}
This is the first dataset for studying EEG-to-video reconstruction \cite{liu2024eegvideo}.
The Laboratory of Brain-like Computing and Machine Intelligence (BCMI) at Shanghai Jiao Tong University (SJTU) acquired 62-channels EEG signals from twenty subjects (10 male and 10 female) during watching natural video clips. These video clips cover 9 coarser natural video classes: \textit{Land Animal, Water Animal, Plant, Exercise, Human, Nutural Scene, Food, Musical Instrument, Transportation}. Each coarser class contains several several classes, composing 40 classes in total. During the experiment, all subjects were instructed to watch a series of color video clips presented with the resolution of 1980 × 1080 (16:9) in full-screen mode on a 25-inch display. Subjects were instructed to watch 7 blocks in total, in each block, the 5 video clips with the same category were displayed continuously, and before playing these 5 same-class videos, there is a hint on the screen to inform the subjects what class they will see next, which will last for three seconds. Consequently, there are 1400 EEG-video pair in total. Following EEG2Video work \cite{liu2024eegvideo}, we use the first 6 blocks as the training set and the last block as the test set.

For EEG features, differential entropy (DE) has been proven to be the most effective handcrafted feature in many field, as it has a balanced ability to discriminate between EEG patterns with low- and high-frequency energy \cite{liu2024eegvideo}. DE feature also compress the raw signals and simplify the computation. We use a 256-point Short-Time Fourier Transform (STFT) with a non-overlapping Hanning window of 1 seconds to calculate the frequency domain features. The DE features are extracted in five frequency bands (delta: 1-4 Hz, theta: 4-8 Hz, alpha: 8-14 Hz, beta: 14-31 Hz, and gamma: 31-49 Hz), which are defined as
\begin{equation}
\begin{aligned}
    h(X)&=-\int_{-\infty}^{+\infty}\frac{1}{\sqrt{2\pi\sigma^2}}e^{-\frac{(x-\mu)^2}{2\sigma^2}}\log(\frac{1}{\sqrt{2\pi\sigma^2}}e^{-\frac{(x-\mu)^2}{2\sigma^2}})dx\\
    &=\frac{1}{2}\log(2\pi e\sigma^2),
\end{aligned}
\end{equation}
where the random variable $X$ obeys the Gaussian distribution $N(\mu,\sigma)$. DE is equivalent to the
logarithmic energy spectrum for a fixed-length EEG sequence in a specific band. Thus, for 62-channel EEG signals, a DE feature sample in five frequency bands has 310 dimensions.

\subsection{fMRI-to-video: CC2017 Dataset}
This is the most widely used benchmark dataset for studying fMRI-to-video reconstruction in both neuroscience and computer science community \cite{wen2018neural, chen2023cinematic, gong2024neuroclips, lu2025animate}. The Laboratory of Integrated Brain Imaging (LIBI) at Purdue University acquired 3T fMRI responses from three subjects during watching natural movies. The movie stimuli contain diverse yet representative of real-life visual experiences, e.g. people in action, moving animals, nature scenes, outdoor or indoor scenes etc. The stimuli include two sets of movie segments: 1) 18 training movie segments, and 2) 5 testing movie segments. The duration of each segment is 8 minutes. During each fMRI scanning session, one segment was presented to the subjects. For each subject, the training movie segments were presented twice and the testing movie segments were presented ten times. In total, there are 11.47 hours of fMRI responses to 3.07 hours of movie stimuli for each subject. We average the 10 testing fMRI segments to formulate one data for inference. Thus there are 8640 training samples and 1200 testing
samples of fMRI-video pairs.

We follow the pre-processing procedure as the NeuroClips \cite{gong2024neuroclips}.
The CC2017 fMRI dataset was preprocessed following a standardized minimal preprocessing approach, as outlined in \cite{wen2018neural}. This pipeline included steps such as artifact removal, motion correction (6 DOF), and spatial normalization to the MNI standard space. Additionally, the fMRI data were mapped onto cortical surfaces, which were aligned with a common surface template. To identify stimulus-activated voxels, we computed voxel-wise correlations between fMRI signals across repetitions of the training movie for each participant. These correlation coefficients were Fisher z-transformed, and the average z scores across 18 movie segments were evaluated using a one-sample t-test. Voxels that showed significant activation (Bonferroni-corrected, P < 0.05) were designated as stimulus-activated and included in further analyses. Specifically, the visual cortex exhibited 13,447, 14,828, and 9,114 activated voxels for the three subjects, respectively. In line with established methodologies \cite{han2019variational, wang2022reconstructing}, we incorporated a 4-second delay in the BOLD signals to account for the hemodynamic response lag when modeling the movie stimulus responses.

\subsection{Data Acquisition}
The open-source datasets used in this paper can be accessed via the following links:

1. SEED-DV: \textcolor{blue}{https://bcmi.sjtu.edu.cn/home/eeg2video/}.

2. CC2017: {\textcolor{blue}{https://purr.purdue.edu/publications/2809/1}}.

\section{Evaluation Metric Implementation}
\label{apd:metrics}
In this section, we detail the metrics we use for the video reconstruction benchmark. The semantic-level metrics to evaluate the quality of generated videos can be roughly classified as frame-based metrics and video-based metrics.

\textbf{Frame-based Metrics} Two levels of metrics are considered to judge the quality of generated frames: the pixel-level and the semantics-level metrics. For the pixel level, we calculate the average structural similarity index measure (SSIM) \cite{wang2004image} of each frame between the ground-truth video and the reconstructed video. For the semantic level, a CLIP-based classifier \cite{radford2021learning} trained on ImageNet \cite{deng2009imagenet} is adopted to compute the \textit{N}-way top-\textit{K} accuracy of predicted frames. If the ground-truth class is within the top-\textit{K} probability of the predicted frames classification results from \textit{N} arbitrary classes (including ground-truth class), the semantic-level reconstruction is regarded successful.

\textbf{Video-based Metric} As the ImageNet classifier is unable to well understand videos, a VideoMAE-based \cite{tong2022videomae} video classifier trained on Kinetics-400 dataset \cite{kay2017kinetics}, which can understand 400 dynamic concepts (e.g., changes, human motions), is applied to compute the video semantic-level accuracy. 

We use the 2-way top 1 (2-I and 2-V) and the 40-way top 1 (40-I and 40-V) to measure the semantic-level accuracy in our paper.
Below, we write an algorithm to clarify how \textit{N}-way top-\textit{K} metric is computed in Algorithm \ref{alg:nwaytopk}.

For the spatiotemporal-level
metrics that measure video consistency, we compute CLIP image embeddings on each frame of the
predicted videos and report the average cosine similarity between all pairs of adjacent video frames,
which is the common metric CLIP-pcc in video editing.

\begin{algorithm}
    \renewcommand{\algorithmicrequire}{\textbf{Input:}}
    \renewcommand{\algorithmicensure}{\textbf{Output:}}
    \caption{\textit{N}-way top-\textit{K} metric}
    \label{alg:nwaytopk}
    \begin{algorithmic}[1]
        \STATE \textbf{Input} pre-trained classifiers $\mathcal{C}_{image}(\cdot)$,  $\mathcal{C}_{video}(\cdot)$, video pair (Generated Video $x$, Corresponding GT Video $\hat{x})$, mode(video-based or image-based)
        \STATE \textbf{Output} success rate $r\in[0,1]$
        \IF{mode=``video-based"}
        \FOR{$100$ trials} 
        \STATE $\hat{y} \leftarrow \mathcal{C}_{video}(\hat{x})$ get the ground-truth class
        \STATE $\{p_0,...,p_{399}\} \leftarrow \mathcal{C}_{video}(x)$ get the output probabilities
        \STATE $\{p_{\hat{y}}, p_{y_1}, ..., p_{y_{n-1}}\} \leftarrow$ pick $n$-1 random classes
        \STATE success if $\mathop{\arg\max}\limits_y\{p_{\hat{y}}, p_{y_1}, ..., p_{y_{n-1}}\} = \hat{y}$
        \ENDFOR
        \STATE $r=$ number of success / $N$
        \ELSE
        \FOR{6 frames}
        \FOR{$100$ trials} 
        \STATE $\hat{y}_i \leftarrow \mathcal{C}_{image}(\hat{x}_i)$ get the ground-truth class
        \STATE $\{p_0,...,p_{999}\} \leftarrow \mathcal{C}_{image}(x_i)$ get the output probabilities
        \STATE $\{p_{\hat{y}_i}, p_{y_{i,1}}, ..., p_{y_{i,n-1}}\} \leftarrow$ pick $n$-1 random classes
        \STATE success if $\mathop{\arg\max}\limits_{y_i}\{p_{\hat{y}_i}, p_{y_{i,1}}, ..., p_{y_{i,n-1}}\} = \hat{y}_i$
        \ENDFOR
        \STATE $r_i=$ number of success / $100$
        \ENDFOR
        \STATE $r=\sum_{i=1}^{6}r_i / 6$ 
        \ENDIF

    \end{algorithmic}  
\end{algorithm}

\section{MindCross Architecture}

MindCross comprises \textit{N} specific encoder and a shared encoder, \textit{N} specific reconstructer and a shared reconstructer, and a shared decoder. The aggregation function is adaptive max pool following the MindBridge \cite{wang2024mindbridge}.
The reconstructer is basically the reverse of encoder. The shared deocder is a 2-layer residual MLP equipped with two linear heads, and the hidden layer size is 2048.
The PyTorch-like architectural code for the encoder and decoder of MindCross is illustrated in Figure \ref{fig:architecture}.

\section{Implementation Details}
\label{apd:implementaion}
MindCross's video generation module is a diffusion-based model called Pyramid-Flow Model \cite{jin2025pyramidal}. We simply used its ``text-to-video" pipeline.
The model embeds each text prompt into two text embeddings, one is called prompt embeddings $p_e$ with a shape of $128\times4096$, another one is called pooled prompt embeddings $p_p$ with a shape of $768$.
We train our model to align the predicted text embeddings $\hat{\textbf{e}}$ with the pooled prompt embeddings $p_p$ for simplifying calculation.
Moreover, we trained another embedding converter $f$ on the ground truth text embeddings for learning the transform $f: p_p \rightarrow p_e$.
After predicting the text embeddings $\hat{\textbf{e}}$, we use the trained converter $f$ to obtain the prompt embeddings $\hat{p_e} = f(\hat{\textbf{e}})$.
We retrieve the CLIP embeddings using $\hat{p_e}$ to offer more dependable embeddings for generating high-quality videos.
For more beautiful videos, we add ``cinematic style, vivid colors" after all text prompts, as suggested in \cite{jin2025pyramidal}.
In this paper, we train our MindCross to align the brain data to the pooled prompt embeddings, and train an MLP to convert pooled prompt embeddings to prompt embeddings.
The guidance scale of the first frame is set to 9.0, and the video guidance scale is set to 5.0.
As the Pyramid-Flow Model can only generate videos with 5:3 ratio, we cropped the central part of the generated videos for the fMRI-to-video benchmarks. For the SEED-DV dataset, whose videos are 16:9, we cropped the gournd truth videos to 5:3 ratio.

The total training loss is balanced by setting the weights of $\alpha, \beta, \gamma$, and $\zeta$ to 0.1. The total calibration loss is balanced by setting the weights of $\alpha'$ and $\beta'$ to 0.1, too. The MindCross models for cross-subject experiments are trained for 1000 epochs, and those for new-subject adaptation is trained for only 200 epochs. Across all experiments, the batch size is set to 256.

For MindBrdige, which is a cross-subject brain-to-image decoding pipeline, we use the same video generation module as our MindCross. So the only difference between our method and MindBridge is the approach of learning text embeddings. For other baseline models, we use their code base on Github. 

1. MinD-Video: {\textcolor{blue}{https://github.com/jqin4749/MindVideo}}.

2. NeuroClips: {\textcolor{blue}{https://github.com/gongzix/NeuroClips}}.

3. Mind-Animator: {\textcolor{blue}{https://github.com/ReedOnePeck/MindAnimator}}.

4. EEG2Video: {\textcolor{blue}{https://github.com/XuanhaoLiu/EEG2Video}}.

5. GLFA: {\textcolor{blue}{https://github.com/chongjg/GLFA-fmri-video}}.

6. MindBridge: {\textcolor{blue}{https://github.com/littlepure2333/MindBridge}}.

\begin{figure*}[h]
\begin{lstlisting}
class ResFuse(nn.Module):
    # input 2 tensors with length of h, output {f(s,r)+r} with length of h
    def __init__(self, h, dropout=0.15):
        super().__init__()
        self.mlp = nn.Sequential(
            nn.Linear(2*h, h),
            nn.LayerNorm(h),
            nn.GELU(),
            nn.Dropout(dropout)
        )
    def forward(self, x, y):
        return self.mlp(torch.concat((x, y), dim=2)) + y



class MindCross(nn.Module):
    def __init__(self, in_dim=310, out_dim_text=77*768, h=2048, subj_list=range(1,N))
        # Adding a new subject position and a shared position
        subj_list = subj_list.append('new')
        subj_list = subj_list.append('shared')
        
        self.encoder = nn.ModuleDict(subj: nn.Sequential(
            nn.Linear(in_dim, h),
            nn.LayerNorm(h),
            nn.GELU(),
            nn.Linear(h, h),
            nn.LayerNorm(h),
            nn.GELU()
        ) for subj in subj_list})

        self.reconstructer = nn.ModuleDict({subj: nn.Sequential(
            nn.Linear(h, in_dim),
            nn.LayerNorm(in_dim),
            nn.GELU(),
            nn.Linear(in_dim, in_dim)
        ) for subj in subj_list})

        self.ResFuse = nn.ModuleDict({str(subj):
            ResFuse(h)
        for subj in subj_list})

        self.shared_decoder = nn.ModuleList([
            nn.Sequential(
                nn.Linear(h, h),
                nn.LayerNorm(h), 
                nn.GELU(),
                nn.Dropout(0.15)
        ) for _ in range(2)])

        self.domain_classifier = nn.Sequential(
            nn.Linear(h, h),
            nn.GELU(),
            nn.Linear(h, self.num_sub)
        )
        
        self.head_text  = nn.Linear(h, out_dim_text)

\end{lstlisting}
\caption{PyTorch-like pseudo code of MindCross architecture.}
\label{fig:architecture}
\end{figure*}

\section{More Experimental Results}

\subsection{New Subject Adaptation}

We show the complete metric in Table \ref{tab:newsub}, where we only select several metric to draw the lineplot in Figure \ref{fig:newsub_line}.

\begin{table*}[htbp]
\setlength\tabcolsep{4.5pt}
\centering
\begin{tabular}{l|lcccccccc}
    \toprule
    ~ & \multirow{3}[3]{*}{Method} & \multirow{3}[3]{*}{\# Data} & \multicolumn{3}{c}{Video-based} & \multicolumn{4}{c}{Frame-based} \\
    \cmidrule(lr){4-6} \cmidrule(lr){7-10}
    ~ & & &  \multicolumn{2}{c}{Semantic-Level} & \multicolumn{1}{c}{ST-Level} & \multicolumn{2}{c}{Semantic-Level} & \multicolumn{2}{c}{Pixel-Level}\\
    \cmidrule(lr){4-5} \cmidrule(lr){6-6} \cmidrule(lr){7-8} \cmidrule(lr){9-10} 
    ~ & & &  2-way    & 40/50-way  & CLIP-pcc  & 2-way   & 40/50-way & SSIM   & PSNR    \\
    \midrule
    \multirow{10}*{\rotatebox{90}{SEED-DV}} & w/o Training Phase & 40    &    0.756\scriptsize{$\pm0.04$}        &        0.096\scriptsize{$\pm0.01$}      &    0.783\scriptsize{$\pm0.64$}      &       0.693\scriptsize{$\pm0.05$}     &   0.085\scriptsize{$\pm0.01$}     &     0.173\scriptsize{$\pm0.08$}     &       8.135\scriptsize{$\pm1.62$}  \\
    ~ & only $\mathcal{L}_{align}$     &    40 &    0.772\scriptsize{$\pm0.02$}        &        0.122\scriptsize{$\pm0.03$}      &    0.741\scriptsize{$\pm0.21$}      &       0.754\scriptsize{$\pm0.03$}     &   0.107\scriptsize{$\pm0.02$}     &     0.184\scriptsize{$\pm0.11$}     &       8.383\scriptsize{$\pm1.53$}  \\
    ~ & MindCross (Ours)     &    40 &    0.785\scriptsize{$\pm0.03$}        &        0.129\scriptsize{$\pm0.03$}      &    0.743\scriptsize{$\pm0.34$}      &       0.741\scriptsize{$\pm0.03$}     &   0.110\scriptsize{$\pm0.03$}     &     0.185\scriptsize{$\pm0.07$}     &       8.437\scriptsize{$\pm1.73$}  \\

    \cmidrule(lr){2-10}
    ~ & w/o Training Phase & 200     &    0.745\scriptsize{$\pm0.05$}        &        0.102\scriptsize{$\pm0.01$}      &    0.715\scriptsize{$\pm0.22$}      &       0.702\scriptsize{$\pm0.06$}     &   0.089\scriptsize{$\pm0.02$}     &     0.179\scriptsize{$\pm0.08$}     &       8.249\scriptsize{$\pm1.46$}  \\
    ~ & only $\mathcal{L}_{align}$     &    200 &    0.778\scriptsize{$\pm0.03$}        &        0.134\scriptsize{$\pm0.03$}      &    0.741\scriptsize{$\pm0.19$}      &       0.739\scriptsize{$\pm0.04$}     &   0.114\scriptsize{$\pm0.01$}     &     0.181\scriptsize{$\pm0.08$}     &       8.533\scriptsize{$\pm1.46$}  \\
    ~ & MindCross (Ours)     &    200 &    0.761\scriptsize{$\pm0.03$}        &        0.137\scriptsize{$\pm0.02$}      &    0.735\scriptsize{$\pm0.39$}      &       0.743\scriptsize{$\pm0.03$}     &   0.117\scriptsize{$\pm0.01$}     &     0.185\scriptsize{$\pm0.06$}     &       8.587\scriptsize{$\pm1.46$}  \\

    \cmidrule(lr){2-10}
    ~ & w/o Training Phase & 600    &    0.749\scriptsize{$\pm0.03$}        &        0.110\scriptsize{$\pm0.03$}      &    0.738\scriptsize{$\pm0.17$}      &       0.721\scriptsize{$\pm0.03$}     &   0.096\scriptsize{$\pm0.02$}     &     0.197\scriptsize{$\pm0.15$}     &       8.361\scriptsize{$\pm1.57$}  \\
    ~ & only $\mathcal{L}_{align}$     &    600 &    0.779\scriptsize{$\pm0.04$}        &        0.141\scriptsize{$\pm0.03$}      &    0.752\scriptsize{$\pm0.37$}      &       0.740\scriptsize{$\pm0.03$}     &   0.122\scriptsize{$\pm0.02$}     &     0.191\scriptsize{$\pm0.09$}     &       8.604\scriptsize{$\pm1.46$}  \\
    ~ & MindCross (Ours)     &    600 &    0.775\scriptsize{$\pm0.03$}        &        0.146\scriptsize{$\pm0.02$}      &    0.750\scriptsize{$\pm0.48$}      &       0.750\scriptsize{$\pm0.02$}     &   0.128\scriptsize{$\pm0.02$}     &     0.189\scriptsize{$\pm0.12$}     &       8.658\scriptsize{$\pm1.39$}  \\

    \midrule
    \multirow{10}*{\rotatebox{90}{CC2017}} & w/o Training Phase & 500   &    0.711\scriptsize{$\pm0.05$}        &        0.104\scriptsize{$\pm0.01$}      &    0.695\scriptsize{$\pm0.15$}      &       0.702\scriptsize{$\pm0.04$}     &   0.097\scriptsize{$\pm0.01$}     &     0.131\scriptsize{$\pm0.11$}     &       8.096\scriptsize{$\pm1.46$}  \\
    ~ & only $\mathcal{L}_{align}$     &    500 &    0.804\scriptsize{$\pm0.04$}        &        0.145\scriptsize{$\pm0.01$}      &    0.724\scriptsize{$\pm0.12$}      &       0.762\scriptsize{$\pm0.04$}     &   0.138\scriptsize{$\pm0.03$}     &     0.145\scriptsize{$\pm0.08$}     &       8.493\scriptsize{$\pm1.46$}  \\
    ~ & MindCross (Ours)     &    500 &    0.817\scriptsize{$\pm0.03$}        &        0.152\scriptsize{$\pm0.02$}      &    0.720\scriptsize{$\pm0.13$}      &       0.767\scriptsize{$\pm0.03$}     &   0.141\scriptsize{$\pm0.01$}     &     0.142\scriptsize{$\pm0.08$}     &   8.551\scriptsize{$\pm1.46$}  \\

    \cmidrule(lr){2-10}
    ~ & w/o Training Phase & 1500    &    0.753\scriptsize{$\pm0.05$}        &        0.132\scriptsize{$\pm0.02$}      &    0.701\scriptsize{$\pm0.33$}      &       0.753\scriptsize{$\pm0.03$}     &   0.126\scriptsize{$\pm0.01$}     &     0.145\scriptsize{$\pm0.04$}     &       8.457\scriptsize{$\pm1.46$}  \\
    ~ & only $\mathcal{L}_{align}$     &    1500 &    0.819\scriptsize{$\pm0.03$}        &        0.173\scriptsize{$\pm0.01$}      &    0.726\scriptsize{$\pm0.44$}      &       0.783\scriptsize{$\pm0.02$}     &   0.160\scriptsize{$\pm0.01$}     &     0.144\scriptsize{$\pm0.07$}     &       8.834\scriptsize{$\pm1.46$}  \\
    ~ & MindCross (Ours)     &    1500 &    0.825\scriptsize{$\pm0.04$}        &        0.178\scriptsize{$\pm0.03$}      &    0.724\scriptsize{$\pm0.24$}      &       0.785\scriptsize{$\pm0.03$}     &   0.165\scriptsize{$\pm0.01$}     &     0.153\scriptsize{$\pm0.03$}     &       8.876\scriptsize{$\pm1.46$}  \\

    \cmidrule(lr){2-10}
    ~ & w/o Training Phase & 4000    &    0.799\scriptsize{$\pm0.03$}        &        0.175\scriptsize{$\pm0.01$}      &    0.726\scriptsize{$\pm0.13$}      &       0.788\scriptsize{$\pm0.08$}     &   0.163\scriptsize{$\pm0.01$}     &     0.147\scriptsize{$\pm0.09$}     &       8.784\scriptsize{$\pm1.46$}  \\
    ~ & only $\mathcal{L}_{align}$     &    4000 &    0.829\scriptsize{$\pm0.03$}        &        0.195\scriptsize{$\pm0.01$}      &    0.735\scriptsize{$\pm0.41$}      &       0.796\scriptsize{$\pm0.03$}     &   0.184\scriptsize{$\pm0.01$}     &     0.165\scriptsize{$\pm0.13$}     &       8.953\scriptsize{$\pm1.46$}  \\
    ~ & MindCross (Ours)     &    4000 &    0.831\scriptsize{$\pm0.03$}        &        0.199\scriptsize{$\pm0.01$}      &    0.734\scriptsize{$\pm0.20$}      &       0.795\scriptsize{$\pm0.05$}     &   0.187\scriptsize{$\pm0.02$}     &     0.163\scriptsize{$\pm0.05$}     &       9.012\scriptsize{$\pm1.46$}  \\
    \bottomrule
\end{tabular}
\caption{The full quantitative results of new subject adaptation in limited data scenario.}
\label{tab:newsub}
\end{table*}

\subsection{Training Phase Loss Function}
Here, we display the whole metric of the ablation study about the loss function in training phase in Table \ref{tab:apd_ab_loss}.

\begin{table*}[]
\setlength\tabcolsep{6pt}
\centering
\begin{tabular}{l|lccccccc}
    \toprule
    ~ & \multirow{3}[3]{*}{Training Loss} & \multicolumn{3}{c}{Video-based} & \multicolumn{4}{c}{Frame-based} \\
    \cmidrule(lr){3-5} \cmidrule(lr){6-9}
    ~ & &  \multicolumn{2}{c}{Semantic-Level} & \multicolumn{1}{c}{ST-Level} & \multicolumn{2}{c}{Semantic-Level} & \multicolumn{2}{c}{Pixel-Level}\\
    \cmidrule(lr){3-4} \cmidrule(lr){5-5} \cmidrule(lr){6-7} \cmidrule(lr){8-9} 
    ~ & &  2-way    & 40/50-way  & CLIP-pcc  & 2-way   & 40/50-way & SSIM   & PSNR    \\
    \midrule
    \multirow{3}*{\footnotesize \rotatebox{90}{EEG}} & $\mathcal{L}_{align}$   &    {0.756\scriptsize{$\pm0.02$}} &  {0.132\scriptsize{$\pm0.04$}}  &    {0.757\scriptsize{$\pm0.57$}}   &       {0.693\scriptsize{$\pm0.03$}}     &   0.128\scriptsize{$\pm0.06$}     &     {0.187\scriptsize{$\pm0.06$}}     &       {8.627\scriptsize{$\pm1.57$}}  \\
    ~ & $ + \mathcal{L}_{da} + \mathcal{L}_{dc}  + \mathcal{L}_{rec}$    &    {0.789\scriptsize{$\pm0.02$}} &  {0.156\scriptsize{$\pm0.03$}}  &    {0.757\scriptsize{$\pm0.57$}}   &       {0.757\scriptsize{$\pm0.03$}}     &   0.125\scriptsize{$\pm0.03$}     &     {0.195\scriptsize{$\pm0.08$}}     &       {8.647\scriptsize{$\pm1.46$}}  \\
    ~ & $ + \mathcal{L}_{diff}$ (Ours)   &    {0.786\scriptsize{$\pm0.02$}} &  {0.154\scriptsize{$\pm0.03$}}  &    {0.758\scriptsize{$\pm0.32$}}   &       {0.752\scriptsize{$\pm0.03$}}     &   0.128\scriptsize{$\pm0.02$}     &     {0.197\scriptsize{$\pm0.06$}}     &       {8.658\scriptsize{$\pm1.43$}}  \\
    \midrule
    \multirow{3}*{\footnotesize  \rotatebox{90}{fMRI}} & $\mathcal{L}_{align}$       & {0.812\scriptsize{$\pm0.02$}} & {0.195\scriptsize{$\pm0.05$}} & {0.758\scriptsize{$\pm0.39$}} & {0.785\scriptsize{$\pm0.03$}} & {0.174\scriptsize{$\pm0.05$}}  & {0.152\scriptsize{$\pm0.07$}} & {8.576\scriptsize{$\pm1.42$}}      \\ 
    ~ & $ + \mathcal{L}_{da} + \mathcal{L}_{dc} + \mathcal{L}_{rec}$    & {0.829\scriptsize{$\pm0.03$}} & {0.213\scriptsize{$\pm0.04$}} & {0.753\scriptsize{$\pm0.41$}} & {0.810\scriptsize{$\pm0.02$}} & {0.194\scriptsize{$\pm0.03$}}  & {0.168\scriptsize{$\pm0.08$}} & {8.957\scriptsize{$\pm1.37$}}      \\ 
    ~ & $ + \mathcal{L}_{diff}$ (Ours)   & {0.830\scriptsize{$\pm0.03$}} & {0.210\scriptsize{$\pm0.03$}} & {0.762\scriptsize{$\pm0.34$}} & {0.802\scriptsize{$\pm0.03$}} & {0.198\scriptsize{$\pm0.03$}}  & {0.163\scriptsize{$\pm0.07$}} & {8.983\scriptsize{$\pm1.54$}}      \\ 
    \bottomrule
\end{tabular}
\caption{Ablation of different training loss.}
\label{tab:apd_ab_loss}
\end{table*}

\subsection{Ablation of Alignment Loss}

Another option to align these shared features' domains is directly minimizing the Kullback–Leibler divergence (KL
divergence) between probabilities produced by the shared
feature representations. We project the shared features using a linear transformation $\delta()$, and caculate the probabilities via a softmax function $\sigma()$:
\begin{equation}
    \mathcal{L}_{da} = \frac{1}{N^2}\sum_{i,k=1}^N \textup{KL}[\sigma(\delta(\textbf{r}^i)), \sigma(\delta(\textbf{r}^k))].
\end{equation}

\noindent Another method to align these shared features' domains is simply minimizing the pairwise \textit{p}-norm:
\begin{equation}
    \mathcal{L}_{da} = \frac{1}{N^2}\sum_{i,k=1}^N \Vert \textbf{s}^i- \textbf{r}^k\Vert_p.
\end{equation}

Here, we display the whole metric of the ablation study about the loss function in calibration phase in Table \ref{tab:apd_ab_loss_align}.

\begin{table*}[]
\setlength\tabcolsep{7pt}
\centering
\begin{tabular}{l|lccccccc}
    \toprule
    ~ & \multirow{3}[3]{*}{Alignment Loss} & \multicolumn{3}{c}{Video-based} & \multicolumn{4}{c}{Frame-based} \\
    \cmidrule(lr){3-5} \cmidrule(lr){6-9}
    ~ & &  \multicolumn{2}{c}{Semantic-Level} & \multicolumn{1}{c}{ST-Level} & \multicolumn{2}{c}{Semantic-Level} & \multicolumn{2}{c}{Pixel-Level}\\
    \cmidrule(lr){3-4} \cmidrule(lr){5-5} \cmidrule(lr){6-7} \cmidrule(lr){8-9} 
    ~ & &  2-way    & 40/50-way  & CLIP-pcc  & 2-way   & 40/50-way & SSIM   & PSNR    \\
    \midrule
    \multirow{3}*{\small \rotatebox{90}{EEG}} & GRL    &    {0.786\scriptsize{$\pm0.02$}} &  {0.154\scriptsize{$\pm0.03$}}  &    {0.758\scriptsize{$\pm0.32$}}   &       {0.752\scriptsize{$\pm0.03$}}     &   0.128\scriptsize{$\pm0.02$}     &     {0.197\scriptsize{$\pm0.06$}}     &       {8.658\scriptsize{$\pm1.43$}}  \\
    ~ & KL     &    0.783\scriptsize{$\pm0.01$}        &        0.150\scriptsize{$\pm0.04$}      &    0.755\scriptsize{$\pm0.52$}      &       0.749\scriptsize{$\pm0.02$}     &   0.132\scriptsize{$\pm0.04$}     &     0.186\scriptsize{$\pm0.04$}     &       8.527\scriptsize{$\pm1.72$}  \\
    ~ & $L_2$ norm   &    0.785\scriptsize{$\pm0.03$}        &        0.149\scriptsize{$\pm0.02$}      &    0.751\scriptsize{$\pm0.48$}      &       0.742\scriptsize{$\pm0.03$}     &   0.125\scriptsize{$\pm0.03$}     &     0.190\scriptsize{$\pm0.17$}     &       8.527\scriptsize{$\pm1.72$}  \\
    \midrule
    \multirow{3}*{\small \rotatebox{90}{fMRI}} & GRL   & {0.830\scriptsize{$\pm0.03$}} & {0.210\scriptsize{$\pm0.03$}} & {0.762\scriptsize{$\pm0.34$}} & {0.802\scriptsize{$\pm0.03$}} & {0.198\scriptsize{$\pm0.03$}}  & {0.163\scriptsize{$\pm0.07$}} & {8.983\scriptsize{$\pm1.54$}}      \\ 
    ~ & KL     &    0.826\scriptsize{$\pm0.03$}        &        0.211\scriptsize{$\pm0.03$}      &    0.756\scriptsize{$\pm0.41$}      &       0.805\scriptsize{$\pm0.03$}     &   0.195\scriptsize{$\pm0.03$}     &     0.158\scriptsize{$\pm0.04$}     &       8.917\scriptsize{$\pm1.49$}  \\
    ~ & $L_2$ norm   &    0.811\scriptsize{$\pm0.04$}        &        0.203\scriptsize{$\pm0.02$}      &    0.747\scriptsize{$\pm0.38$}      &       0.792\scriptsize{$\pm0.05$}     &   0.200\scriptsize{$\pm0.08$}     &     0.157\scriptsize{$\pm0.03$}     &       8.867\scriptsize{$\pm1.37$}  \\
    \bottomrule
\end{tabular}
\caption{Ablation of different alignment loss functions.}
\label{tab:apd_ab_loss_align}
\end{table*}

\subsection{Additional Visualization}
\begin{figure}[H]
  \centering
    \subfigure[]{
       \centering
       \includegraphics[width=0.22 \textwidth]{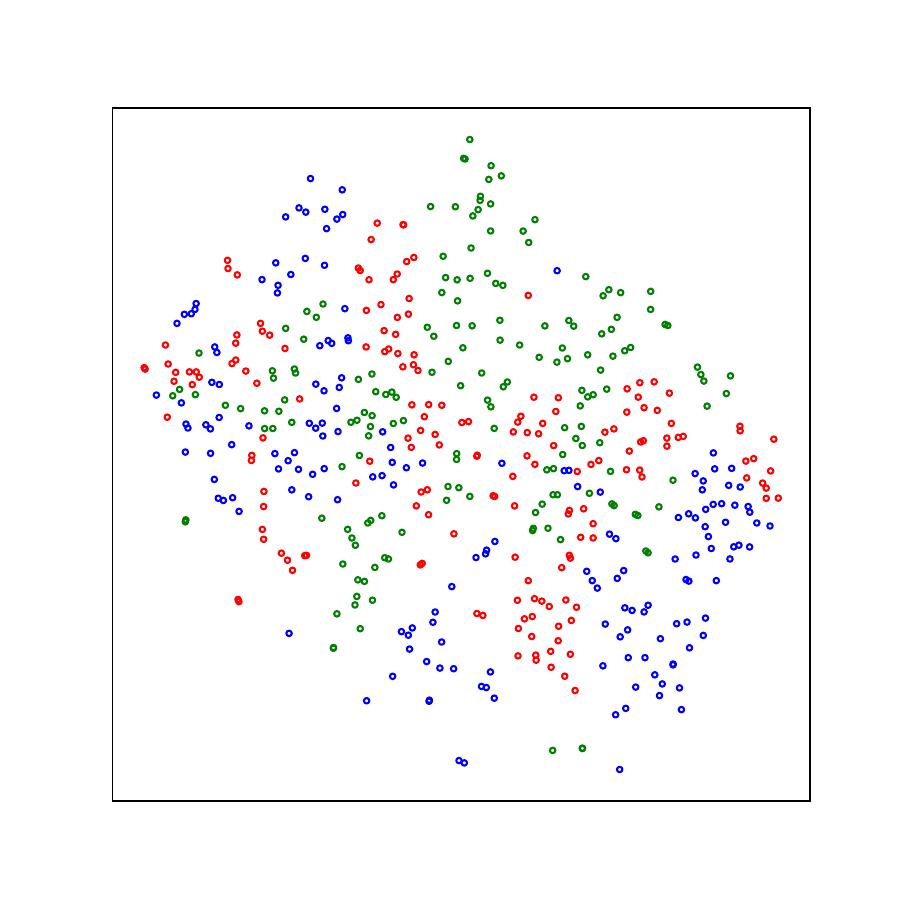}
    }
    \subfigure[]{
        \centering
        \includegraphics[width=0.22 \textwidth]{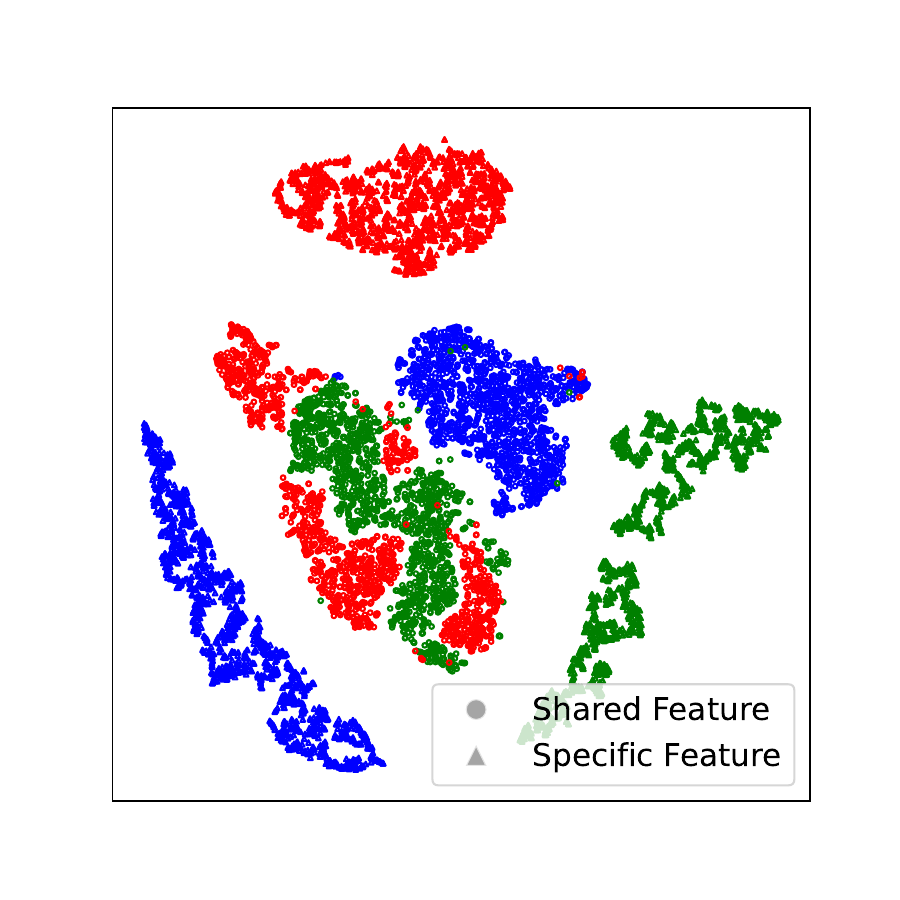}
    }
  \caption{
  Feature visualization of MindCross. (a) Original brain data. The color represents different subjects. (b) Learned features of MindCross. The triangle stand for the specific features and circle stand for the shared features.
  }
  \label{fig:shaspecfeature_apd_cc2017}
\end{figure}

In Figure \ref{fig:shaspecfeature}, we plot the feature learned by MindCross on the SEED-DV dataset. Here, we present the feature visualization on the CC2017 dataset in Figure \ref{fig:shaspecfeature_apd_cc2017}.
For the original data, we only select 2\% out of all training dataset (due to the time limitation), and using an nn.AdaptiveAvgPool1d(2048) layer to crop all fMRI data to a constant length for tSNE visualization.
But for the learned feature, we selected a bit more data than the original data.
It can be seen that the subject-related and subject-invariant information are well learned by our MindCross, indicating the versatality of our MindCross on both fMRI data and EEG data. 

\subsection{Additional Decoding Results}

\begin{figure*}
    \centering
    \includegraphics[width=0.9\linewidth]{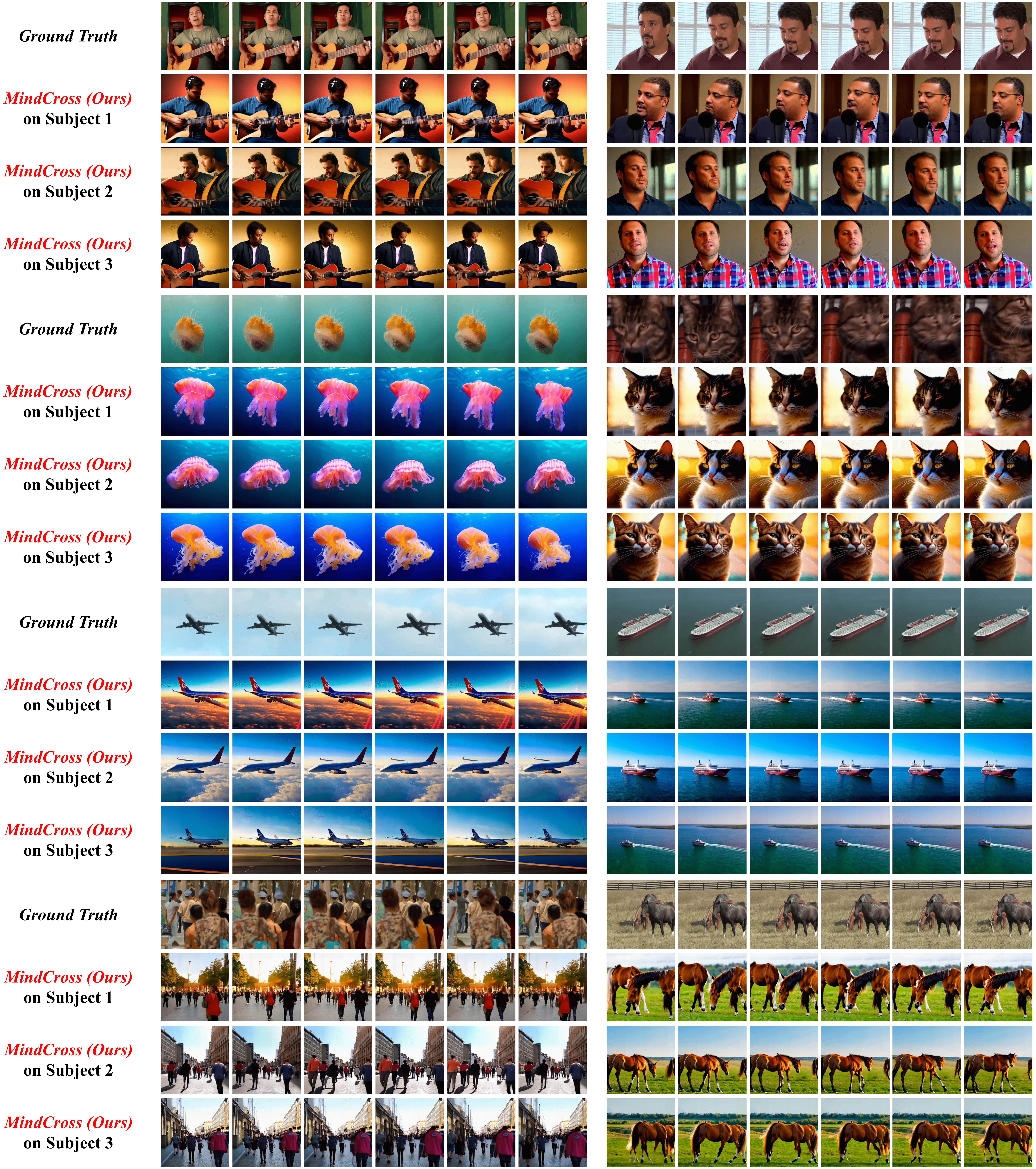}
    \caption{Additional reconstruction samples on the CC2017 dataset.}
    \label{fig:apd_more_cc}
\end{figure*}

\begin{figure*}
    \centering
    \includegraphics[width=0.8\linewidth]{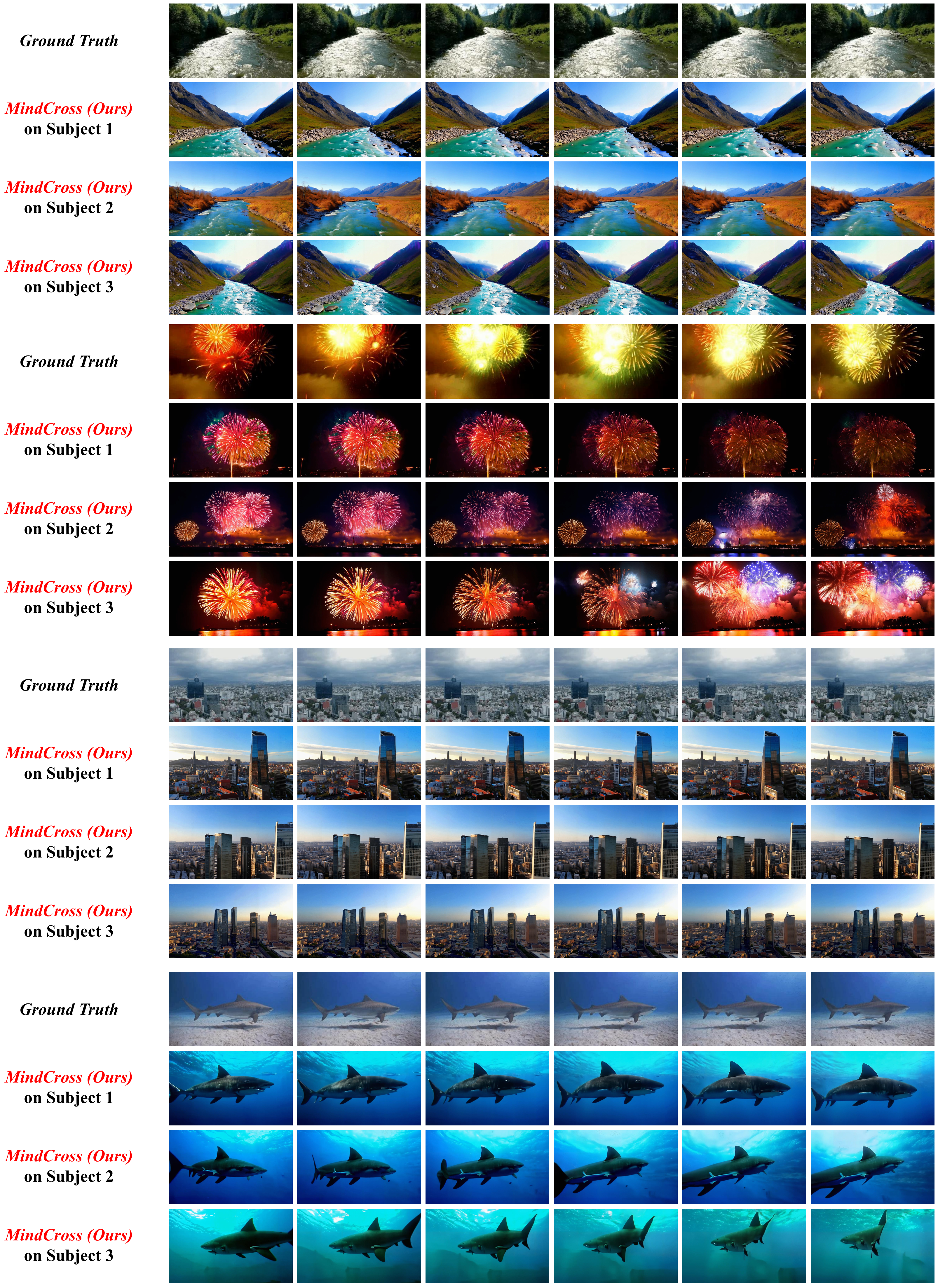}
    \caption{Additional reconstruction samples on the SEED-DV dataset.}
    \label{fig:apd_more_seeddv}
\end{figure*}

Figure \ref{fig:apd_more_cc} and Figure \ref{fig:apd_more_seeddv} show additional decoded videos from different subjects in the study. These findings demonstrate that our proposed model is able to decode high-quality videos with precise semantics across different subjects. Furthermore, the rich collection of videos shown below highlights the model’s ability to decode a wide range of semantic content in human brain activity. This includes videos related to objects, human motion, scenes, animals, and more.

\section{Broad Impacts}

Our research can be used for visualize our mind across different subjects, offering a novel approach for listening the inner world of people patients with mental illnesses like autistic and depression. However, every coin has two sides. Personal privacy may leak through our brain activities and be abused by malicious attackers for reading one’s mind from EEG signals without acknowledgment. More strict regulations are supposed to be made for protecting the privacy of people’s biological data by government and medical institutions.

\end{document}